\documentclass[manuscript]{aastex}

\slugcomment{Not to appear in Nonlearned J., 45.}

\shorttitle{The Resonant Nuclear Reaction}
\shortauthors{Liu et al.}

\begin{document}

\title{Resonant nuclear reaction $^{23}$Mg $(p,\gamma)$ $^{24}$Al in strongly screening magnetized
neutron star crust}

\author{Jing-Jing Liu\altaffilmark{1}, and Dong-Mei, Liu\altaffilmark{1}}
\affil{$^{1}$ College of Marine Science and Technology, Hainan
Tropical Ocean University, Sanya, Hainan 572022, China}

\altaffiltext{1}{Corresponding author: syjjliu68@qzu.edu.cn}

\begin{abstract}
Based on the relativistic theory of superstrong magnetic field
(SMF), by using three models of Lai (LD), Fushiki (FGP), and ours
(LJ), we investigate the influence of SMFs due to strong electron
screening (SES) on the nuclear reaction $^{23}$Mg $(p, \gamma)$
$^{24}$Al in magnetars. At relatively low density environment (e.g.,
$\rho_7<0.01$) and $1<B_{12}<10^2$, our screening rates are in good
agreement with those of LD and FGP. However, in relatively high
magnetic fields (e.g., $B_{12}>10^2$), our reaction rates can be
1.58 times and about three orders of magnitude larger than those of
FGP and LD, respectively ($B_{12}$, $\rho_7$ are in units of
$10^{12}$G, $10^7\rm{g~cm^{-3}}$). The significant increase of
strongly screening rates can imply that more $^{23}$Mg will escape
from the Ne-Na cycle due to SES in a SMF. As a consequence, the next
reaction $^{24}$Al $(\beta^+, \nu)$ $^{24}$Mg will produce more
$^{24}$Mg to participate in the Mg-Al cycle. Thus, it may lead to
synthesize a large amount of production of $A > 20$ nuclides in
magnetars, respectively.

\end{abstract}


\section{Introduction}

In the dense sites of universe, such as novae, X-ray bursts and
supernova, there are explosive hydrogen burning process in high
temperature  and high hydrogen environments. This burning is called
the rapid-proton (rp) process \citep{Wallace81}. In the stage of
hydrogen burning, the proton capture reactions and
$\beta^{+}$-decays (rp-process) will be ignited in the nuclei whose
mass numbers $A>20$. For example, the timescale of the proton
capture reaction of $^{23}$Mg in the Ne-Na cycle at sufficient high
temperature is shorter than that of the $\beta^{+}$-decay.
Therefore, some $^{23}$Mg will kindle and escape from the Ne-Na
cycle by proton capture. The $^{23}$Mg leaks from the Ne-Na cycle
into the Mg-Al cycle synthesizing a large amount of heavy nuclei.
Thus the reaction $^{23}$Mg $(p,\gamma)$ $^{24}$Al in stellar
environment is an important reaction for producing heavy nuclei.
\citet{Wallace81} firstly discussed the reaction rate of $^{23}$Mg
$(p, \gamma)$ $^{24}$Al. Then, \citet{Iliadis01} also investigated
this nuclear reaction rate. \citet{Kubono95} reconsidered the rate
by considering four resonances and the structure of $^{24}$Al. Based
on some new experimental information on $^{24}$Al excitation
energies, \citet{Herndl98, Visser07}, and \citet{Lotay08} carried
out an estimation of the rate. However, they all seem to have
overlooked the influence of electron screening on nuclear reaction.

In the pre-supernova stellar evolution and nucleosynthesis, the
strong electron screening (SES) is always a challenging and
interesting problem. Some works \citep{Bahcall02, Liu13, Liu14,
Liu16, Liu17a, Liu17b} have been done on stellar weak-interaction
rates and thermonuclear reaction rates. In the high-density
surrounding, some SES models have been widely investigated, such as
Salpeter model \citep{Salpeter54, Salpeter69}, Graboske model
\citep{Graboske73}, and Dewitt model \citep{Dewitt76}. Recently
these issues were discussed by \citet{Liolios00, Liolios01},
\citet{Kravchuk14}, and \citet{Liu13}. However, they neglected the
effects of SES on thermonuclear reaction rate in superstrong
magnetic field (SMF).

It is widely known that nuclear reaction rates at low energies play
a key role in energy generation in stars and the stellar
nucleosynthesis. The bare reaction rates are modified in stars by
the screening effects of free and bound electrons. The knowledge of
the bare nuclear reaction rates at low energies is important not
only for the understanding of various astrophysical nuclear
problems, but also for assessing the effects of host material in low
energy nuclear fusion reactions in matter.

It is universally accepted that the surface dipole magnetic field
strengths of magnetars are in a range from $10^{13}$ to $10^{15}$G
\citep{Peng07, Gao11, Gao13, Gao15, Gao17a, Gao17b, Li16, Lai01}.
The momentum space of the electron gas is modified substantially by
so intense magnetic fields. The electron Fermi energy and nuclear
reaction are also affected greatly by a SMF in magnetars.

Anamalou x-ray pulsars (AXPs) and soft gamma-ray repeaters (SGRs)
are conceived as magnetars, which are a kind of special pulsars
powered by their magnetic energy \citep{Duncan92}. The Fermi energy
of the electrons will increase with magnetic field and quantum
effects of electron gas will be very obvious in a SMF. As we all
know, the positive energy levels of electrons must abide by Landau
quantization. The distribution of the electron in the momentum space
will be strongly modified by a SMF. Some authors discussed this
issue in detail in strong magnetic fields of magnetars. For
instance, \citet{Gao15, Gao17a, Gao17b} investigated not only the
spin-down and magnetic field evolutions, but also the electron
Landau level effects on emission properties of magnetars.

In this paper, according to the relativistic theory in a SMF
\citep{Peng07, Gao11, Gao13, Gao15, Gao17b}, we discuss the problem
of SES and then investigate the effect of SES on the thermonuclear
reaction within three different models (i.e., our model (LJ), Lai
model (LD)\citep{Lai91, Lai01}, and Fushiki model
(FGP)\citep{Fushiki89}) on the surface of magnetars.

Our work differs from previous work of \citet{Liu16} about the
discussion of nuclear reaction rates. Firstly, in Liu (2016), though
we cited several work from Gao et al., but it is not familiar to the
calculations involved in electron Fermi energy in a superhigh
magnetic field, a non-relativistic electron cyclotron solution was
applied when calculating the rates. Secondly, our previous work
\citet{Liu16} did not give a comparison among LJ, LD, and FGP models
in the case with a SMF. Finally, we analyze the nuclear reaction
rates in a SMF and also give a comparison for our model with Dewitt
model\citep{Dewitt76}, and Liolios model\citep{Liolios00}, in which
the SMF were not taken into consideration. Maybe SES universally
occur in pulsars, and the screening rate calculations in a SMF is of
importance to the future studies on cooling, nucleosynthesis,and
emission properties of magnetars.

In this paper, following the works of \citet{Peng07}, and
\citet{Gao11, Gao13, Gao15, Gao17b}, we calculate the resonant
reaction rates in the case with SMF and without SMF in several
screening models. In the case of the former, the results from LD and
FGP models will be compared with those of our model, while in the
latter case, the results from Dewitti and Liolio models also will be
compared. We derive new results for SES theory and the screening
rates for nuclear reaction in relativistic strong magnetic fields.

The article is organized as follows. In the next Section, we analyse
three SES models in a SMF of magnetars. In Section 3 we discuss the
effects of SES on the proton capture reaction rate of $^{23}$Mg, in
which the four resonances contributions will also be considered. The
results and discussions will be shown in Section 4. The article is
closed with some conclusions in Section 5.

\section{The SES in SMF}

In astrophysical systems, the SMF may have significant influence on
the quantum processes. In this Section, we will study three models
of the electron screening potential (ESP) in SMF, i.e., LJ model, LD
model, and FGP model.

\subsection{ESP in our model}

The rate of nuclear reaction in high density matter is affected by
the fact that the clouds of the electrons surrounding nuclei alter
the interactions among nuclei. The positive energy levels of
electrons in SMF are given by \citep{Landau77}
\begin{equation}
   \frac{\varepsilon_n}{m_{\rm{e}} c^2}=[(\frac{p_{\rm{z}}}{m_{\rm{e}}c})+1+2(n+\frac{1}{2}+\sigma)b]^{1/2}=(p^2_{\rm{z}}+\Theta)^{1/2},
\label{1}
\end{equation}
 where $\Theta=1+2(n+\frac{1}{2}+\sigma)b$, $n=0,1,2,3....$,
 $b=\frac{B}{B_{\rm{cr}}}=0.02266B_{12}$, $B_{12}$
 is the magnetic fields in units of $10^{12}$G, i.e.,
 $B_{12}\equiv B/10^{12}$G,
$B_{\rm{cr}}=\frac{m^2_{\rm{e}} c^3}{e\hbar}=4.414\times10^3$G is
the electron quantum critical magnetic field, and $p_{\rm{z}}$ is
the electron momentum along the field, $\sigma$
 is the spin quantum number of an electron, when $n=0$,$\sigma=1/2$, and when $n\geq1$, $\sigma=\pm1/2$.

In an extremely strong magnetic field $(B\gg B_{\rm{cr}})$, the
Landau column becomes a very long and narrow cylinder along the
magnetic field. According to the Pauli exclusion principle, the
electron number density should be equal to its microscopic state
density. By introducing the electron Landau level stability
coefficient, the Fermi energy of the electron is given by
\citep{Gao13,zhu16}
\begin{eqnarray}
 U_{\rm{F}}&=&5.91\times10^4(\frac{B}{B_{{\rm{cr}}}})^{1/6}(\frac{\rho Y_{\rm{e}}}{\rho_0\times0.00564})^{1/3}\nonumber\\
 &=&5.91\times10^4(\frac{B}{B_{{\rm{cr}}}})^{1/6}(\frac{n_{\rm{e}}}{0.00564\times\rho_0
 N_{\rm{A}}})^{1/3}\rm{keV},
\label{4}
\end{eqnarray}
where $\rho_0=2.8\times 10^{14} {\rm g/cm}^3$  is the standard
nuclear density.

In order to evaluate the Thomas-Fermi screening wave-number
$K_{{\rm{TF}}}^{{\rm{LJ}}}$, we defined a parameter
$D^{{\rm{LJ}}}(U_{\rm{e}})$ and according to Eq.(3), we have
\begin{equation}
n_e=0.00564\rho_0 N_A(\frac{U_{\rm{F}}}{5.91\times 10^4b^{1/6}})^3
\label{6}
\end{equation}
\begin{eqnarray}
 D^{{\rm{LJ}}}(U_{\rm{F}})&=&\frac{\partial n_{\rm{e}}}{\partial U_{\rm{F}}}=\frac{\partial}{\partial U_{\rm{F}}}(0.00564\rho_0 N_A(\frac{U_{\rm{F}}}{5.910\times 10^4b^{1/6}})^3)\nonumber\\
 &=&4.9913\times10^{7}n_{\rm{e}}^{2/3}b^{-1/6}~~~\rm{cm^{-3}~KeV^{-1}}.
 \label{6}
\end{eqnarray}

According to Eq.(5), the Thomas-Fermi screening wave-number
$K_{{\rm{TF}}}^{{\rm{LJ}}}$ is given by \citep{Ashcroft76}
\begin{eqnarray}
(K_{{\rm{TF}}}^{{\rm{LJ}}})^2&=&4\pi e^2 D^{{\rm{LJ}}}(U_{\rm{F}})
 =4\pi e^2\frac{\partial n_{\rm{e}}}{\partial U_{\rm{F}}}\nonumber\\
 &=&6.269\times10^{7}e^2(n_e)^{2/3}b^{-1/6}~~~\rm{cm^{-3}}.
 \label{8}
\end{eqnarray}

By using the uniform electron gas model \citep{Kadomtsev71}, the
binding energy of the magnetized condensed matter at zero pressure
can be estimated. The energy per cell can be written as
\begin{equation}
E_{\rm{total}}=E_{\rm{k}}+E_{\rm{latt}}=\frac{3\pi^2e^2z_{j}^3}{8b_1^2r_{i}^6}+\frac{9e^2z^{\frac{5}{3}}}{10r_e}~~~\rm{MeV},
 \label{9}
\end{equation}
where the first term is the kinetic energy and the second term is
the lattice energy. $r_{i}=z^{1/3}r_{\rm{e}} a_0$ is the
Wigner-Seitz cell radius, $a_0=0.529\times10^{-8}$cm is the Bohr
radius, and $r_{\rm{e}}=(3/4\pi n_{\rm{e}})^{1/3}$ is the mean
electron spacing. $z_{j}$ is the charge number of the species $j$.
$b_1=B/B_0=425.4B_{12}=1.9773\times10^4b$ and
$B_0=m_{\rm{e}}^2ce^3/\hbar^3=2.3505\times10^{-9}$G  is  the natural
(atomic) unit for the field strength \citep{Lai01}. For the
zero-pressure condensed matter, we require $dE_{\rm{total}}/dr_i=0$,
so we have
\begin{equation}
r_i=r_{i0}=0.0371z_j^{1/5}b^{-2/5}a_0~~~\rm{cm}. \label{10}
\end{equation}

By using linear response theory, the energy correction per cell due
to non-uniformity is given by \citep{Lattimer85}
\begin{eqnarray}
E_{{\rm{TF}}}^{{\rm{LJ}}}(r_i,z_j)&=&-\frac{18}{175}(K_{{\rm{TF}}}^{{\rm{LJ}}}r_i)^2\frac{(z_je)^2}{r_i}\nonumber\\
&=&-\frac{1.30\times10^{-6}e^6(n_{\rm{e}})^{4/3}z_{j}^{9/5}}{b^{11/15}}~~~\rm{MeV}.
\label{11}
\end{eqnarray}

For the relativistic electrons, the influence from exchange free
energy were discussed by Refs.\cite{Stolzmann96, Yakovlev89}. Their
works showed that the correlation correction is very small.
Therefore, in this paper we have neglected the correction of Coulomb
exchange free energy interaction in the electron gas model. By
taking into consideration of the Coulomb energy and Thomas-Fermi
correction due to non-uniformity of the electron gas, the energy per
cell should be corrected as
\begin{equation}
 E_{{\rm{s}}}^{{\rm{LJ}}}(r_{i},z_{j})=E_{{\rm{k}}}(r_{i},z_{j})-U_{{\rm{coul}}}(r_{i},z_{j})-E_{{\rm{TF}}}^{{\rm{LJ}}}(r_{i},z_{j}).
\label{12}
\end{equation}

For two interaction nuclides, the energy required to bring two
nuclei with nuclear charge numbers $z_1$ and $z_2$ so close together
that they essentially coincide differs from the bare Coulomb energy
by an amount which in the Wigner-Seitz approximation is
\begin{equation}
U_{{\rm{sc}}}=E_{{\rm{s}}}(r_{i},z_{12})-E_{{\rm{s}}}(r_{i},z_1)-E_{\rm{s}}(r_{i},z_2),
\label{13}
\end{equation}
where $z_{12}=z_{1}+z_{2}$. If the electron distribution is rigid,
the contribution to from $E_{\rm{s}}$ the bulk electron energy
cancel in expression (11), and the screening potential is simply
given as
\begin{eqnarray}
 U_{\rm{sc}}&=&E_{\rm{coul}}(r_{i},z_{12})-E_{\rm{coul}}(r_{i},z_1)-E_{\rm{coul}}(r_{i},z_2)\nonumber\\
 &=&6.5984\times10^4b^{2/5}(z_{12}^{9/5}-z_{1}^{9/5}-z_{2}^{9/5})
 \rm{MeV},
\label{14}
\end{eqnarray}
where we assume the electron density is uniform, and the screening
potential is independent of the magnetic field.

From expression (9), the change of the screening potential due to
the compressibility of the electrons in the zero-pressure magnetized
condensed matter can obtained as
\begin{eqnarray}
 \delta E_{\rm{TF}}^{\rm{LJ}}&=&-\frac{18}{175}(K_{\rm{TF}}^{\rm{LJ}}r_{i})^2\frac{e^2(z_{12}^2-z_1^2-z_2^2)}{r_{i}}\nonumber\\
&=&-\frac{1.30\times10^{-6}e^6n_{\rm{e}}^{4/3}(z_{12}^{9/5}-z_1^{9/5}-z_2^{9/5})}{b^{11/15}}.
\label{15}
\end{eqnarray}

In accordance with the above discussions, the total screening
potential is the sum of the screening potential with a uniformity
distribution and a corrected screening potential with a
non-uniformity distribution. The screening potential in SMF is given
by
\begin{equation}
U_{\rm{sc}}^{\rm{LJ}}=U_{\rm{sc}}+\delta E_{\rm{TF}}^{\rm{LJ}}.
\label{16}
\end{equation}

\subsection{ESP in LD model}
\citet{Lai01} and \citet{Lai91} discussed the equation of state and
the electron energy in a SMF. In a SMF the electron number density
$n_{\rm{e}}$ is related to the chemical potential $U_{\rm{e}}$ by
\begin{eqnarray}
 n_{\rm{e}}=\frac{1}{(2\pi\widehat{\rho})^2\hbar}\sum_0^\infty g_{n0} \int_{-\infty} ^{+\infty} f dp_{\rm{z}}\nonumber\\
 =\frac{1}{(2\pi\widehat{\rho})^2\hbar}\sum_0^\infty g_{n0}\int_{-\infty}^{+\infty}[1+\exp(\frac{E-U_{\rm{e}}}{kT})]^{-1}dp_{\rm{z}},
\label{17}
\end{eqnarray}
where $\widehat{\rho}=(\hbar
c/eB)^{1/2}=2.5656\times10^{-10}B_{12}^{1/2}\rm{cm}$ is the electron
cyclotron radius (the characteristic size of the wave packet), and
$E=[c^2p_z^2+m_ec^4(1+nb)]^{1/2}$ is the free electron energy,
$g_{n}$ is the spin degeneracy of the Landau level, $g_{00}=1$ and
$g_{n0}=2$ for $n\geqslant 1$, and the Fermi-Dirac distribution is
given by
\begin{equation}
f=[1+\exp(\frac{E-U_{\rm{e}}}{kT})]^{-1}. \label{19}
\end{equation}

The electron Fermi energy including the electron rest mass is given
by
\begin{equation}
 n_{\rm{e}}=\frac{1}{2\pi^{3/2}\lambda_{\rm{Te}}\widehat{\rho}^2}\sum_{(n=0)}^{\infty} g_{n}I_{-1/2}(\frac{U_{\rm{e}}-n\hbar\omega_{\rm{ce}}}{kT}),
\label{20}
\end{equation}
where the thermal wavelength of the electron is
$\lambda_{\rm{Te}}=(2\pi\hbar^2/m_{\rm{e}} kT)^2$, and the Fermi
integral is written as
\begin{equation}
I_{n}(y)=\int_0^\infty \frac{x^{n}}{\exp(x-y)+1}dx. \label{21}
\end{equation}

The binding energy of the magnetized condensed matter at zero
pressure can be estimated using the uniform electron gas model.
Under the condition of super-strong magnetic field, the Fermi energy
$U_{\rm{F}}$ is less than the cyclotron energy
$\hbar\omega_{\rm{ce}}$, the electrons only occupy the ground Landau
level. According to their viewpoint of \citep{Lai01}, the
Thomas-Fermi screening wave-number is given by
\begin{equation}
 (K_{\rm{TF}}^{\rm{LD}})^2=4\pi e^2D^{\rm{LD}}(\varepsilon_{\rm{F}})=4\pi e^2 \frac{\partial n_{\rm{e}}}{\partial \varepsilon_{\rm{F}}}=4\pi e^2 \frac{\partial
n_{\rm{e}}}{\partial U_{\rm{F}}}, \label{22}
\end{equation}
where $\partial n_{\rm{e}}/\partial \varepsilon_{\rm{F}}$ is the
density of states per unit volume at the Fermi surface.
$\varepsilon_{\rm{F}}=P_{\rm{F}}^2/2m_{\rm{e}}$. From Eq.(6.16) of
\citet{Lai01}, so we have
\begin{equation}
 D^{\rm{LD}}=\frac{\partial n_{\rm{e}}}{\partial \varepsilon_{\rm{F}}}=\frac{3.79\times10^6b^2r_{\rm{e}}^3}{e^2}.
\label{23}
\end{equation}

The Thomas-Fermi screening wave-number will be given by
\begin{equation}
K_{\rm{TF}}^{\rm{LD}}=(\frac{4}{3\pi^2})^{1/2}b_1r_{\rm{e}}^{3/2}=6.901\times10^3br_{\rm{e}}^{3/2}.
\label{24}
\end{equation}

Using the linear response theory, the energy correction (in atomic
units) per cell due to non-uniformity can be calculated and gives by
\citep{Lai01}
\begin{equation}
E_{\rm{TF}}^{\rm{LD}}(r_{i},z_{j})=-\frac{18}{175}(K_{\rm{TF}}^{\rm{LD}}r_{i})^2\frac{e^2z_{j}^2}{r_{i}}=-0.0139b_1^2r_{i}^4z_{j}.
\label{25}
\end{equation}

The uniform electron gas model can be improved by taking into
consideration of the Coulomb energy and Thomas-Fermi correction due
to non-uniformity of the electron gas. When the electron density is
assumed to be uniform, the screening potential is independent of the
magnetic field. The change of the screening potential due to the
compressibility of the electrons for the zero-pressure magnetized
condensed matter can obtained
\begin{equation}
 \delta
 E_{\rm{TF}}^{\rm{LD}}=-2.5236\times10^{-4}b^{2/5}(z_{12}^{9/5}-z_{1}^{9/5}-z_{2}^{9/5}).
\label{26}
\end{equation}

When we summed of a screening potential with a uniformity
distribution and a corrected screening potential with a
non-uniformity distribution, the screening potential in a SMF is
given by
\begin{equation}
U_{\rm{s}}^{\rm{LD}}=U_{\rm{sc}}+\delta E_{\rm{TF}}^{\rm{LD}}.
\label{27}
\end{equation}

\subsection{ESP in FGP model}
The influence of SES in a SMF on nuclear reaction was also discussed
in detail by \citet{Fushiki89} (hereafter FGP). The electron Coulomb
energy by an amount which in the Wigner-Seitz approximation in a SMF
was given by
\begin{equation}
U_{\rm{sc}}^{\rm{FGP}}=E_{\rm{atm}}(r_{i},z_{12})-E_{\rm{atm}}(r_{i},z_1)-E_{\rm{atm}}(r_{i},z_2),
\label{28}
\end{equation}
where $E_{\rm{atm}}(r_{i}, z_{j})$ is the total energy of
Wigner-Seitz cell. If the electron distribution is rigid, the
contribution to $E_{\rm{atm}}(r_{i}, z_{j})$ from the bulk electron
energy cancel, the electron screening potential at high density can
be expressed as
\begin{equation}
U_{\rm{sc}}^{\rm{FGP}}=E_{\rm{latt}}(r_{i},z_{12})-E_{\rm{latt}}(r_{i},z_1)-E_{\rm{latt}}(r_{i},z_2),
\label{29}
\end{equation}
where $E_{\rm{latt}}(r_i, z_j)$ is the electrostatic
 energy of Wigner-Seitz cell and $E_{atm}(r_{i}, z_{j})=-0.9z_{j}^{5/3}e^2/r_{\rm{e}}$.
Due to the influence of the compressibility of the electron, the
change in the screening potential is given by \citep{Fushiki89}
\begin{eqnarray}
 \delta U_{\rm{s}}^{\rm{FGP}}=-\frac{54}{175}(\frac{e^2}{r_{\rm{e}}})\frac{1}{n_{\rm{e}}}\frac{\partial n_{\rm{e}}}{\partial U_{\rm{e}}}[(z_{12})^{7/3}-(z_{1})^{7/3}-(z_{2})^{7/3}]\nonumber \\
 =-\frac{54}{175}(\frac{e^2}{r_{\rm{e}}})\frac{1}{n_{\rm{e}}}D^{\rm{FGP}}[(z_{12})^{7/3}-(z_{1})^{7/3}-(z_{2})^{7/3}],
\label{30}
\end{eqnarray}
where
\begin{equation}
 D^{\rm{FGP}}=823.1481\frac{r_{\rm{e}}n_{\rm{e}}}{e^2}(\overline{\frac{A}{z}})^{4/3}\rho^{-4/3}B_{12}^2.
\label{31}
\end{equation}

The Thomas-Fermi screening wave-number will be given by
\begin{equation}
 (K_{\rm{TF}}^{\rm{FGP}})^2=1.0344\times 10^4 r_{\rm{e}}n_{\rm{e}}(\overline{\frac{A}{z}})^{4/3}\rho^{-4/3}B_{12}^2.
\label{32}
\end{equation}

Thus, the corresponding result for the changes in the screening
potential in a SMF is
\begin{eqnarray}
 \delta U_{\rm{s}}^{\rm{FGP}}=-0.254(\overline{\frac{A}{z}})^{4/3}\rho^{-4/3}B_{12}^2 [(z_{12})^{7/3}-(z_{1})^{7/3}-(z_{2})^{7/3}] \nonumber \\
 =-494.668(\overline{\frac{A}{z}})^{4/3}\rho^{-4/3}b^2[(z_{12})^{7/3}-(z_{1})^{7/3}-(z_{2})^{7/3}]\rm{MeV},
\label{33}
\end{eqnarray}
where $(\overline{A/z})$ is the average ratio of $A/z$ , which
corresponding to the mean molecular weigh per electron. Thus the
electron screening potential in a SMF of FGP model is given by
\begin{equation}
U_{\rm{s}}^{\rm{FGP}}=U_{\rm{sc}}+\delta E_{\rm{TF}}^{\rm{FGP}}
=U_{\rm{sc}}+\delta U_{\rm{s}}^{\rm{FGP}}. \label{34}
\end{equation}

\section{Resonant reaction process and rates}
 \subsection{Calculations of resonant reaction rates with and without SES}
The reaction rates are summed of contribution from the resonant
reaction and non-resonant reaction. In the case of a narrow
resonance, the resonant cross section $\sigma_r$ is approximated by
a Breit-Wigner expression \citep{Fowler67}
\begin{equation}
 \sigma_{\rm{r}}(E)=\frac{\pi\omega}{\kappa^2}\frac{\Lambda_{i}(E)\Lambda_{f}(E)}{(E-E_{\rm{r}}^2)+\frac{\Lambda^2_{\rm{total}}(E)}{4}},
\label{35}
\end{equation}
where $\kappa$ is the wave number, the entrance and exit channel
partial widths are $\Lambda_{i}(E)$ and $\Lambda_f(E)$ ,
respectively. $\Lambda_{\rm{total}}(E)$ is the total width, and the
statistical factor, $\omega$ is given by
\begin{equation}
 \omega=(1+\delta_{12})\frac{2J+1}{(2J_1+1)(2J_2+1)},
\label{36}
\end{equation}
where the spins of the interacting nuclei and the resonance are
$J_1$, and $J_2$, respectively, $\delta_{12}$ is the Kronecker
symbol.

The partial widths is dependent on the energy, and can be written
as\citep{Lane58}
\begin{equation}
 \Lambda_{i, f}=2\vartheta_{i,f}^2\psi_{l}(E,a)=\Lambda_{i,f}\frac{\psi_{l}(E,a)}{\psi_{l}(E_{f},a)}.
\label{37}
\end{equation}

The penetration factor $\psi_{l}$ is associated with $l$ and $a$,
which are the relative angular momentum and the channel radius,
respectively. $a=1.4(A_1^{1/3}+A_2^{1/3})$ fm. $\Lambda_{i,f}$ is
the partial energy widths at the resonance process. $E_{\rm{r}}$ and
$\vartheta_{i,f}^2$ is the reduced widths, given by
\begin{equation}
 \vartheta_{i,f}^2=0.01\vartheta_{\rm{w}}^2=\frac{0.03\hbar^2}{2Aa^2}.
\label{38}
\end{equation}

Based on the above analysis, in the phases of explosive stellar
burning, the narrow resonance reaction rates without SES are
determined by \citep{Schatz98, Herndl98}
\begin{eqnarray}
 \lambda_{\rm{r}}^0=N_{\rm{A}}\langle{\sigma v}\rangle_{\rm{r}}=1.54\times10^{11}(AT_9)^{-3/2}\nonumber\\
 \times \sum_{i} \omega \gamma_{i}\exp(-11.605E_{r_{i}}/T_9)~~~\rm{cm^3 mol^{-1} s^{-1}},
\label{39}
\end{eqnarray}
where $N_{\rm{A}}$ is Avogadro's constant, $A$ is the reduced mass
of the two collision partners, $E_{r_i}$ is the resonance energies
and $T_9$ is the temperature in unit of $10^9$ K. The $\omega
\gamma_{i}$ is the strength of resonance in units of MeV and given
by
\begin{equation}
 \omega \gamma_{i}=(1+\delta_{12})\frac{2J+1}{(2J_1+1)(2J_2+1)}\frac{\Lambda_{i}\Lambda_{f}}{\Lambda_{\rm{total}}}.
\label{40}
\end{equation}

On the other hand, due to SES the reaction rates of narrow resonance
is given by
\begin{eqnarray}
 \lambda_{\rm{r}}^s&=&F_{\rm{r}} N_{\rm{A}}\langle{\sigma v}\rangle_{\rm{{r'}}} \nonumber\\
 &=&1.54\times10^{11}(AT_9)^{-3/2}\sum_{i} \omega \gamma_{i}\exp(-11.605 E^{'}_{r_i}/T_9) \nonumber\\
 &=&1.54\times10^{11}F_{\rm{r}} (AT_9)^{-3/2}\nonumber\\
 &&\times \sum_{i} \omega \gamma_{i} \exp(-11.605 E_{r_i}/T_9)  ~\rm{cm^3 mol^{-1} s^{-1}},
\label{41}
\end{eqnarray}
where $F_{\rm{r}}$ is the screening enhancement factor (hereafter
SEF). The values of $E^{'}_{r_i}$ should be measured by experiment,
but it is too hard to provide sufficient data. In general and
approximate analysis, we have
$E^{'}_{r_i}=E_{r_i}-U_0=E_{r_i}-U_{\rm{s}}$.

\subsection{The screening model of resonant reaction rates in the
case without SMF}

\subsubsection{Dewitt model}

 \citet{Dewitt76} discussed the problem of
thermonuclear ion-electron screening at some densities. Based on a
statistical mechanical theory for the screening function, the
influence of the electron screening on the nuclear reaction process
also was investigated in their paper. The strong electron screening
potential function is given by \citep{Dewitt76}
\begin{eqnarray}
 H_{12}^{\rm{sc}}&=&\frac{e^2}{r_{\rm{e}}kT}\{0.9(\overline{z})^{1/3}(z_{12}^{5/3}-z_{1}^{5/3}-z_{2}^{5/3})\nonumber\\
 &&+c_1(\overline{z})^{2/3}(z_{12}^{4/3}-z_{1}^{4/3}-z_{2}^{4/3})\}\nonumber\\
 &&+[c_2(\overline{z})^{-2/3}(z_{12}^{2/3}-z_{1}^{2/3}-z_{2}^{2/3})],
\label{42}
\end{eqnarray}
 where $c_1=0.2843$ and $c_2=0.4600$, and the $\overline{z}$, the average charge of ionic, is given by
\begin{equation}
 \overline{z}=\sum_i z_{i}f_{i}=\sum_{\rm{i}}z_{i}\frac{n_i}{n_{\rm{I}}},
\label{43}
\end{equation}
where $n_{i}$ and $n_{\rm{I}}$ are the ion densities of nuclear
species $i$ and I of the total system, respectively.

The screening enhancement factor (hereafter SEF) in Dewitt model is
written as
\begin{equation}
 F_r^0(\rm{Dew})=\exp(H_{12}^{\rm{sc}}).
\label{44}
\end{equation}

\subsubsection{Liolios model}

At astrophysical energies the electron-screening acceleration in
laboratory fusion reactions always play a key role and is an
interesting problem for astrophysics. Based on a mean-field model,
\citet{Liolios00} studied the screened nuclear reactions at
astrophysical energies. The electron screening potential in Liolios
screened Coulomb model is given as \citep{Liolios00}
\begin{equation}
 U_0^{\rm{Lios}}=\frac{15}{8}\frac{z_1z_2e^2}{\Xi},
\label{45}
\end{equation}
where
\begin{equation}
 \Xi=(\frac{15}{8\pi
 z_{i}^2})^{1/3}a_0=0.8853a_0(z_1^{2/3}+z_2^{2/3})^{1/2},
\label{46}
\end{equation}
The SEF for the resonant reaction in Liolios model is
\begin{equation}
 F_{\rm{r}}^0(\rm{Lios})=\exp(\frac{11.605U_0^{Lios}}{T_9}).
\label{47}
\end{equation}

\subsection{The screening model of resonant reaction rates in SMFs}

In this Subsection, we will discuss the screening potential in the
strong screening limit. The dimensionless parameter $(\Gamma)$,
which determines whether or not correlations between two species of
nuclei $(z_1, z_2)$ are important, is given by
\begin{equation}
\Gamma=\frac{z_1z_2e^2}{(z_1^{1/3}+z_2^{1/3})r_{\rm{e}}kT},
\label{48}
\end{equation}
Under the conditions of $\Gamma\gg1$, the nuclear reaction rates
will be influenced appreciably by SES. According to the above three
SES models (LD, FGP, LJ) in SMFs, the three enhancement factors for
resonant reaction process in SMFs can be expressed as follows
\begin{equation}
F_{\rm{r}}^{\rm{B}}(\rm{LD})=\exp(\frac{11.605U_{\rm{s}}^{\rm{LD}}}{T_9}),
\label{49}
\end{equation}
\begin{equation}
F_{\rm{r}}^{\rm{B}}(\rm{FGP})=\exp(\frac{11.605U_{\rm{s}}^{\rm{FGP}}}{T_9}),
\label{50}
\end{equation}
\begin{equation}
F_{\rm{r}}^{\rm{B}}(\rm{LJ})=\exp(\frac{11.605U_{\rm{s}}^{\rm{LJ}}}{T_9}).
\label{51}
\end{equation}

\section{Numerical results}

\subsection{Analysis of the results on a SEF}

The strong magnetic fields modify significantly the properties of
the matter and always play a critical role in astronomical
conditions. Figure 1 presents the variations of ESP as a function of
$B_{12}$ for our SES model. The SMF has only a slight influence on
ESP when $B_{12}>3\times10^3$ and $\rho_7<1$. But the ESP increases
greatly when $B_{12}<1.4 \times10^3$ and $\rho_7<1$ ($B_{12}$,
$\rho_7$ are in units of $10^{12}$G, $10^7\rm{g~cm^{-3}}$,
respectively). Numerical results in our model show that the maximum
value of ESP reaches to $0.1$ MeV. Figure 2 (a) presents the ESP in
LD model as a function of $B_{12}$. The ESP increases rapidly and
reaches the maximum value of $0.008442$ MeV at $B_{12}=80$, then
decreases with increasing of a SMF.

Based on the Thomas-Fermi and Thomas-Fermi-Dirac approximations,
\citet{Fushiki89} analyzed the electron Fermi energy, electron
Landau level, and SES problem in a SMF. The results show that, as a
consequence of the field dependence of the screening potential,
magnetic fields can significantly increase nuclear reaction rates
\citep{Fushiki89}. According to electron screening model of
Ref.\citep{Fushiki89} (hereafter FGP model ) in a SMF, Figure 2 (b)
shows the ESP as a function of $B_{12}$ under some typical
astrophysical conditions. The ESP increases greatly when
$B_{12}<10^3$ and gets to the maximum value of $0.0188$ MeV at
$B_{12}=580.7$ and $\rho_7=0.1$. Then the ESP decreases around two
orders of magnitude when $10^3<B_{12}<2\times10^3$ at $\rho_7=0.1$.

The influence of SES in a SMFs on nuclear reaction is mainly
reflected by the SEF. We discuss the influence of SES on SEF by
three models (LD, FGP, LJ) from Figure 3 to Figure 4. One finds that
the SEF of LD model is a sensitive parameter for a SMF and
temperature. The maximum value of a SEF is about $1.632$ for
$B_{12}=78.17$ and $T_9=0.2$, as shown in Figure 3, where $T_9$ is
the temperature in units of $10^9$K. But for $B_{12}>219.3$ the SEF
is less than $1.001$. Figure 4 presents the SEF as function of
$B_{12}$ of FGP and LJ models. From sub-figures 4(a) and 4 (b), one
find that the shifty trend of SEF in FGP model is in good agreement
with those of LD at low density (e.g. $\rho_7=0.01$). The maximum
value of SEF of FGP model is about $1.66$ for $B_{12}=84.18$ and
$\rho_7=0.01$. On the contrary, the SEF increases with increasing of
$B_{12}$ at relatively high density (e.g. $\rho_7=1$), then gets to
the maximum value of $5.166$ at $T_9=0.2$. Sub-figures 4(c) and 4(d)
show that in LJ model show that the SEF increases with increasing of
$B_{12}$, and the maximum value will reach up to $5.056$ for
$B_{12}=1000$, $T_9=0.2$ and $\rho_7=0.01$.

In Figure 5, some comparisons of the resonant SEF are shown among
the models of LJ, LD, and FGP for typical astronomical conditions in
a SMF. The results of LD model are well agreement with those of FGP
for relatively low density (e.g., $\rho_7\leqslant 0.01$).
Nevertheless, the SEF of our model decreases placidly with the
increasing of $B_{12}$ and $T_9$ to compare with those of LD and
FGP.

The SES problem always plays important roles in stellar evolution
process. Based on a statistical mechanical theory for the screening
function, \citet{Dewitt76} investigated the influence of the
electron screening on nuclear reaction. Based on a mean-field model,
\citet{ Liolios00} also studied the effect about screened nuclear
reactions. However, they neglected the influence of SMFs on SES. We
compare the SEF of the two models (Dewitt, and Liolios model) with
those of LD, FGP, and LJ. One can conclude that the SEF of Dewitt
model is larger than those of other three SES models for
$B_{12}<140$, $\rho_7=0.01$ and $T_9<0.17$, shown as in Figure 6.
However, when $T_9<0.18$, $\rho_7=0.01$, the results of our model
are larger than those of Dewitt and Liolios. At a relatively high
density (e.g., $\rho_7=0.1$), the SEFs of LD, FGP and LJ models
decrease due to SMFs and is lower than those of Dewitt model. The
results obtained by \cite{Dewitt76} amount to an overestimation of
the screening effect because of their neglect of spatial dependence
of the screening function.

Table 1 shows some information of SEF for the five typical models at
some astronomical conditions. The results of LD, FGP, and LJ are
always lower than those of Liolios and Dewitt due to a SMF. The SEF
of our model decreases very greatly with increasing of density and
temperature when $B_{12}=10^3$. It is because that the ESP increases
very rapidly as SMF increases. The higher the ESP, the larger the
influence on SES becomes. On the contrary, the SEF of LD decreases
with increasing of magnetic fields because ESP is reduced. The SEF
of FGP model gets to the maximum of $1.929$ when $B_{12}=10^3,
\rho_7=1, T_9=0.5$ and then decreases slowly as the density and
temperature increase.

The Thomas-Fermi screening wave-number $K_{\rm{TF}}$ is a very key
parameter, which strongly depends on the electron number density and
ESP. In consequence the electron number density and ESP will play
important roles in a SMF. \citet{Lai91}, analyzed in detail the
electron Fermi energy and electron number density in a SMF based on
the works of \citet{Canuto68, Canuto71, Kubo65}, and
\citet{Pathria03}. By using the uniform electron gas model and
linear response theory, \citet{Lai01}discussed the electron energy
(per cell) corrections due to non-uniformity in a SMF. According to
their theory, we study the ESP and the SES model (i.e., LD model).
The results show that the ESP decreases as the magnetic fields
increase due to the diminution of electron chemical potential. The
LD model is valid only in the condition of $K_{\rm{TF}} r_i\ll1$ at
lower densities, because they investigated the non-uniformity effect
only through detailed electronic (band) structure calculations.

The electron chemical potential is a pivotal parameter, which is
closely related to the electron number density and exchange energy.
Based on Thomas-Fermi-Dirac approximation, it is given as
\citep{Fushiki89}
\begin{equation}
 U_{\rm{F}}=U_{\rm{e}}=\frac{\partial w_{\rm{ex}}}{\partial n_{\rm{e}}}=\frac{r_{\rm{cyc}}}{\pi
 a_0}\hbar\omega_0nI(n),\label{56}
\end{equation}
where $w_{\rm{ex}}$ is the exchange energy and $I(n)$ can be found
in Ref. \citep{Fushiki89}. By using the linear response theory,
\citet{Fushiki89} discussed the exchange energy and electron
chemical potential in the lowest Landau level for non-uniformity
electron gas in a SMF. They analyzed the SES problem in a SMF and
their results shown that a SMF only the lowest Landau level is
occupied by electrons on the condition of
$r_{\rm{e}}>(3\pi/8)^{1/3}r_{\rm{cyc}}$ or equivalently
$\rho<7.04\times10^3B_{12}^{3/2}$(A/z)$\rm{g/cm^3}$. The cyclotron
radius in the lowest Landau level orbital is give by
$r_{\rm{cyc}}=(2\hbar c/eB)^{1/2}\simeq
3.36\times10^{-10}B_{12}^{-1/2}$. FGP used the expression of
$n_{\rm{e}}\partial n_{\rm{e}}/\partial
U_{\rm{F}}=(3/2)n_{\rm{e}}/U_{\rm{F}}$ in dealing with $\partial
n_{\rm{e}}/\partial U_{\rm{F}}$. In FGP model, they thought at high
density the exchange correction is very small, thus they neglected
the exchange correction to $\partial n_{\rm{e}}/\partial U_{\rm{F}}$
and had $n_{\rm{e}}\partial n_{\rm{e}}/\partial
U_{\rm{F}}=(1/2)n_{\rm{e}}/U_{\rm{F}}$ in a SMF. Due to different
ways of dealing with exchange correction under this condition, the
SEF of FGP model has some difference compared with other SES models.

According to statistical physics the microscopic state number
$dxdydzdp_{x}dp_{y}dp_{z}$ can be given by
$dxdydzdp_{x}dp_{y}dp_{z}/h^{3}$ in a 6-dimension phase-space. The
number of states occupied by completely degenerate relativistic
electrons per volume is calculated by \citep{Canuto68, Canuto71}
\begin{eqnarray}
 &N_{\rm{phase}}= \sum_{p_{x}}\sum_{p_{y}}\sum_{p_{z}}
 = \frac{1}{h^{3}}\int_{-\infty}^{\infty}\int_{-\infty}^{\infty}\int_{-\infty}^{\infty}dp_{x}dp_{y}dp_{z}\nonumber\\
 &= \frac{1}{h^{3}}\int_{0}^{p_{\rm F}}dp_{z}\int_{0}^{\infty}p_{\bot}dp_{\bot}\int_{0}^{2\pi}d\theta
 = \frac{\pi p_{\rm F}}{h^{3}}\int_{0}^{\infty}dp_{\bot}^{2},
\label{54}
 \end{eqnarray}
where $\theta~=~tan^{-1}p_{y}/p_{x}$, $p_{\bot}^{2}\rightarrow
m^{2}c^{4}\frac{B}{B_{\rm cr}}2n$, So,
$\int_{0}^{\infty}dp_{\bot}^{2}\rightarrow
\sum_{n=0}^{\infty}\omega_{n}$, and the $\omega_{n}$ is the
degeneracy of the $n$-th electron Landau level in relativistic
magnetic field, and can be calculated by \citep{Canuto71, Kubo65,
Pathria03}
\begin{eqnarray}
\omega_{n}&=& \frac{1}{h^{2}}\int_{0}^{2\pi}d\phi\int_{k_{1} < p_{\bot}^{2}< k_{2}}p_{\bot}dp_{\bot}= \frac{2\pi}{h^{2}}\frac{(k_{2}-k_{1})}{2}\nonumber\\
&=&\frac{1}{2\pi}(\frac{\hbar}{m_{e}c})^{-2}\frac{B}{B_{\rm
cr}}=\frac{b}{2\pi}(\frac{\hbar}{m_{e}c})^{-2} ~~, \label{55}
\end{eqnarray}
where $k_{1}= 2nm_e^{2}c^{2}\frac{B}{B_{\rm cr}}= 2nbm_e^{2}c^{2}$,
and $k_{2}= 2(n+1)bm_e^{2}c^{2}$.

Based on the works of \citet{Peng07, Gao13}, which introduced the
Dirac $\delta$-function and considered Pauli exclusion principle, we
discuss the SES problem in a SMF. Our results show that the stronger
the magnetic field, the higher Fermi energy of electrons becomes.
The ESP increases with SMF and the maximum value of ESP is 0.1 MeV
in a SMF. The SEF also increases greatly and its maximum approaches
to 5.0 MeV (e.g. $\rho_7=0.01, T_9=0.2, B_{12}=10^3$G).

\subsection{Investigation of the nuclear reaction rates}

In the explosive hydrogen burning stellar environments, the nuclear
reaction $^{23}$Mg$(p,\gamma)$$^{24}$Al plays a key role because of
breaking out the Ne-Na cycle to heavy nuclear species (i.e., Mg-Al
cycle). Therefore, it is very important to accurate determinate the
rates for the reaction $^{23}$Mg$(p, \gamma)$$^{24}$Al. However, the
resonance energy has a large uncertainty due to the inconsistent
$^{24}$Mg($^3$He,t)$^{24}$Al measurements mentioned. So it may lead
to a factor of 5 variation in the reaction rate at $T_9=0.25$
because of its exponential dependence on $E_{\rm{r}}$
\citep{Visser07}. Some authors discussed the contributions from
several important resonance states, such as \citep{Wallace81,
Wiescher86, Kubono95, Visser07}. In order to reduce the uncertainty
of the reaction rates in this paper, we reference some information
about this reaction and the values of the $E_{\rm{{r_{i}}}},
E_{\rm{x}}$ and corresponding to $\omega \gamma_{i}$ and some
average values of $\omega \gamma_{i}$ are adopted and listed in
Table 2. According to these information, we analysis the total rates
for these five SES models.

\begin{table*}[htb]\footnotesize
 \centering
 \begin{minipage}{140mm}
  \caption{The comparisons of the resonant SEFs for Dewitt, Liolios, LD,  FGP and LJ models
  in several typical astronnomical conditions. The former two models are in the case without
  SES and SMFs, while the latter three models are in the case with SES and SMFs.}
  \begin{tabular}{@{}rrrrrrrrrrr@{}}
  \hline
 &  & & &\multicolumn{3}{c}{$B_{12}=10$}& &\multicolumn{3}{c}{$B_{12}=10^3$}\\

\cline{5-7} \cline{9-11} \\
 $\rho_7$ &$T_9$& $F_{\rm{r}}^0$(Lios)& $F_{\rm{r}}^0$(Dew)& $F_{\rm{r}}^{\rm{B}}$(LD)& $F_{\rm{r}}^{\rm{B}}$(FGP)
 &$F_{\rm{r}}^{\rm{B}}$(LJ) & & $F_{\rm{r}}^{\rm{B}}$(LD)& $F_{\rm{r}}^{\rm{B}}$(FGP)&$F_{\rm{r}}^{\rm{B}}$(LJ)\\

 \hline
 0.01 &0.1 &1.7475  &3.8973  &1.6956  &1.6964  &0.1749  &&1.0725e-15 &1.3472e-13  &25.5680\\
 0.05 &0.1 &1.7475  &10.9451 &1.6956  &1.7013  &8.4513e-4   &&1.0725e-15  &0.6045  &19.5717\\
 0.1  &0.2 &1.3219  &4.3605  &1.3021  &1.3045  &0.0022  &&3.2750e-8   &2.4894  &3.8848\\
 0.1  &0.3 &1.2051  &2.5873  &1.1922  &1.1941  &0.0174       &&1.0221e-5  &1.8371  &2.4713\\
 0.2  &0.3 &1.2045  &3.3934  &1.1924  &1.1939  &9.1604e-4   &&1.0236e-5   &2.4990  &2.1361\\
 0.3  &0.4 &1.1497  &2.8170  &1.1411  &1.1422  &7.7822e-4 &&1.8097e-4 &2.1178  &1.6055\\
 1.0  &0.5 &1.1181  &3.5124  &1.1113  &1.1122  &6.2630e-7     &&0.0011    &1.9290  &0.9512\\
 10   &0.7 &1.0830  &7.2692  &1.0780  &1.0791  &6.2012e-9     &&0.0072    &1.6142  &0.0894\\
\hline
\end{tabular}
\end{minipage}
\end{table*}


\begin{table*}[htb]\footnotesize
\centering
\begin{minipage}{140mm}
\caption{Resonance parameters for the reaction $^{23}$Mg
$(p,\gamma)$ $^{24}$ Al.} \label{t.lbl}

\begin{tabular}{ccrrrrrrl}
\hline

 $E_{\rm{x}}$ (MeV)  \footnote{is adopted from Ref. \citep{Endt98}} &$E_{\rm{x}}$ (MeV) \footnote{from Ref.\citep{Visser07}}
 &$J^{\pi}$ & $E_{\rm{{r_{i}}}}$ (MeV) \footnote{ from Ref.\citep{Audi95}} & $\Gamma_{\rm{p}}$ &
 $\Gamma_{\gamma}$ & $\omega \gamma_{i}$(meV) \footnote{ from Ref.\citep{Herndl98}}
 &$\omega \gamma_{i}$(meV) \footnote{from Ref.\citep{Wiescher86}} & $\omega \gamma_{i}$(meV) \footnote{is adopted in this paper}\\
 \hline
2.349$\pm$0.020 &2.346$\pm$0.000 &$3^{+}$  &0.478   &185   &33  &25  &27  &26\\
2.534$\pm$0.013  &2.524$\pm$0.002 &$4^{+}$  &0.663   &2.5e3 &53  &58  &130 &94\\
2.810$\pm$0.020  &2.792$\pm$0.004 &$2^{+}$  &0.939   &9.5e5 &83  &52  &11  &31.5\\
2.900$\pm$0.020  &2.874$\pm$0.002 &$3^{+}$  &1.029   &3.4e4 &14  &12  &16  &14\\

\hline
\end{tabular}

\end{minipage}
\end{table*}
\begin{table*}[htb]\footnotesize
\centering \caption{Comparisons of the rates of $\lambda^0_{r}$,
which are in the case without SES with those of the
 LD ($\lambda^{\rm{scB}}_{\rm{r}}(\rm{LD})$), FGP ($\lambda^{\rm{scB}}_{\rm{r}}(\rm{FGP})$) and our calculations
$\lambda^{\rm{scB}}_{\rm{r}}(\rm{LJ})$ in the case with SES for some
typical
 astronomical conditions at $B_{12}=10$, respectively.
$S_i=\lambda_{\rm{r} {i}}^{\rm{scB}}/\lambda_{\rm{r}}^0$, $i=1, 2,
3$ denote the rates of LD, FGP, and LJ model, respectively.}
\begin{minipage}{140mm}
\begin{tabular}{lllllllll}
\hline \hline
\multicolumn{5}{r}{$B_{12}=10$} \\
\cline{4-6}
$\rho_7$ &$T_9$ & $\lambda^0_{\rm{r}}$ &$\lambda^{\rm{scB}}_{\rm{r}}(\rm{LD})$  &$\lambda^{\rm{scB}}_{\rm{r}}(\rm{FGP})$ &$\lambda^{\rm{scB}}_{\rm{r}}(\rm{LJ})$ &$S_{1}$ &$S_2$ &$S_3$\\
\hline
 0.01 &0.1 &1.0942e-19   &1.8552e-19   &1.8561e-19   &1.9138e-20  &1.6956  &1.6964  &0.1749\\
 0.02&0.1&1.0942e-19&1.8552e-19&1.8598e-19&4.0569e-21&1.6956&1.6998&0.0371\\
 0.03&0.1&1.0942e-19&1.8552e-19&1.8608e-19&1.0413e-21&1.6956&1.7007&0.0095\\
 0.03 &0.2 &4.2967e-8   &5.5949e-8   &5.6034e-8   &4.1916e-9   &1.3021  &1.3041  &0.0976\\
 0.04&0.2&4.2967e-8&5.5949e-8&5.6041e-8&2.2448e-9&1.3021&1.3043&0.0522\\
 0.05 &0.2 &4.2967e-8   &5.5949e-8   &5.6044e-8   &1.2491e-9   &1.3021  &1.3043  &0.0291\\
 0.1 &0.2 &4.2967e-8   &5.5949e-8   &5.6051e-8   &9.3383e-11  &1.3021  &1.3045  &0.0022\\
 0.2 &0.4 &0.0163  &0.0186  &0.0186  &8.5713e-5   &1.1411  &1.1422  &0.0053\\
 0.3 &0.5 &0.1925  &0.2140  &0.2141  &6.2713e-4   &1.1114  &1.1122  &0.0033\\
 0.5 &0.6 &0.9764  &1.0663  &1.0669  &8.5404e-4   &1.0920  &1.0927  &8.7465e-4\\
 0.7 &0.8 &7.2550  &7.7500  &7.7537  &0.0079  &1.0682  &1.0687  &0.0011\\
 1.0 &0.9 &14.0604 &14.9100 &14.9162 &0.0050  &1.0604  &1.0609  &3.5785e-4\\

\hline
\end{tabular}
\end{minipage}
\end{table*}

\begin{table*}[htb]\footnotesize
\centering \caption{Comparisons of the rates of
$\lambda^0_{\rm{r}}$, which are in the case without SES and SMFs
with those of the
 LD ($\lambda^{\rm{scB}}_{\rm{r}}(\rm{LD})$), FGP ($\lambda^{\rm{scB}}_{\rm{r}}(\rm{FGP})$) and our calculations
$\lambda^{\rm{scB}}_{\rm{r}}(\rm{LJ})$ in the case with SES for some
typical astronomical conditions at $B_{12}=10^3$, respectively.
 The $S_i$ is the same as in Table 3.}
\begin{minipage}{140mm}
\begin{tabular}{lllllllll}
\hline \hline
\multicolumn{5}{r}{$B_{12}=10^{3}$} \\
\cline{4-6}
$\rho_7$ &$T_9$ & $\lambda^0_{\rm{r}}$ &$\lambda^{\rm{scB}}_{\rm{r}}(\rm{LD})$  &$\lambda^{\rm{scB}}_{\rm{r}}(\rm{FGP})$ &$\lambda^{\rm{scB}}_{\rm{r}}(\rm{LJ})$ &$S_{1}$ &$S_2$ &$S_3$\\
\hline

 0.01 &0.1 &1.0942e-19   &1.1735e-34  &1.4740e-32   &2.7975e-18   &1.0725e-15   &1.3472e-13  &25.5680\\
 0.02 &0.1 &1.0942e-19 &1.1735e-34 &6.4612e-24  &2.5883e-18 &1.0725e-15 &5.9052e-5 &23.6555\\
 0.03 &0.1 &1.0942e-19 &1.1735e-34 &1.5306e-21  &2.4177e-18 &1.0725e-15 &0.0140 & 22.0969\\
 0.03 &0.2 &4.2967e-8   &1.4072e-15  &5.0819e-9   &2.0198e-7   &3.2750e-8   &0.1183 &4.7007\\
 0.04 &0.2 &4.2967e-8 &1.4072e-15 &1.7120e-8   &1.9575e-7  &3.2750e-8 &0.3984 &4.5559\\
 0.05 &0.2 &4.2967e-8   &1.4072e-15  &3.3408e-8   &1.9009e-7   &3.2750e-8   &0.7775 &4.4240\\
 0.1  &0.2 &4.2967e-8   &1.4072e-15  &1.0696e-7   &1.6692e-7   &3.2750e-8   &2.4894  &3.8848\\
 0.2  &0.4 &0.0163  &2.9459e-6   &0.0324  &0.0288  &1.8097e-4   &1.9876  &1.7669\\
 0.3  &0.5 &0.1925  &1.9523e-4 &0.3509  &0.2812  &0.0010  &1.8227  &1.4604\\
 0.5  &0.6 &0.9764  &0.0031  &1.6579  &1.1949  &0.0032  &1.6979  &1.2237\\
 0.7  &0.8 &7.2550  &0.0976  &10.8789 &7.8159  &0.0135  &1.4995  &1.0773\\
 1.0  &0.9 &14.0604 &0.3053  &20.2526 &13.6742 &0.0217  &1.4404  &0.9725\\

\hline
\end{tabular}
\end{minipage}
\end{table*}

Tables 3 and 4 give a brief description of the factor $S_i$ $ (i=1,
2, 3)$ for LD, FGP, and LJ models when $B_{12}=10, 10^3$,
respectively. As the density and temperature increase, the results
of LD model are in good agreement with those of FGP, but
disagreement with our results at $B_{12}=10$. This is because that
the electron Fermi energy of our model is lower than those of LD and
FGP in relatively low magnetic fields. As the magnetic fields
increase from $B_{12}=10$ to $10^3$, the factor $S_3$ increases
about $2\sim3$ orders magnitude (i.e., from 0.1749 to 25.5680 and
from 0.0022 to 3.8848) when $\rho_7=0.01, T_9=0.1$ and $\rho_7=0.1,
T_9=0.2$, respectively. When $B_{12}=10^3$ the factor $S_3$ is about
39.74, 5.69, 1.56 times larger than $S_2$ (FGP model) at
$\rho_7=0.03, T_9=0.2$, $\rho_7=0.05, T_9=0.2$ and $\rho_7=0.1,
T_9=0.2$, respectively. From what has been discussed above, the LD
model maybe only adapts to the relatively low magnetic field and low
density surroundings. The FGP and LD models are both unadapted to
relatively low density, and high magnetic field surroundings (e.g.
$\rho_7<0.1, B_{12}>10^2$). However, our model can be well adapted
to relatively high magnetic field and low density surroundings (e.g.
$B_{12}>10^2, \rho_7<0.05$).

Summing up the above discussions, our calculations show that this
SES effect in a SMF can increase nuclear reaction rates of $^{23}$Mg
$(p, \gamma)$$^{24}$Al by several orders magnitude. A more precise
thermonuclear rates of $^{23}$Mg $(p, \gamma)$$^{24}$Al will help us
to constrain the determination of nuclear flow out of the Ne-Na
cycle, and production of $A\geq20$ nuclides, in explosive hydrogen
burning over a temperature range of $0.2\leq T \leq 1.0$ GK.

\section{Conclusions}

In this paper, based on the relativistic theory in a SMF, we
investigate the problem of SES, and the SES influence on the nuclear
reaction of $^{23}$Mg $(p, \gamma)$$^{24}$Al by LD, FGP, and LJ
strong screening models in a SMF. The results show that the SES
thermonuclear reaction rates have a remarkable increase in a SMF.
The rates can increase by around three orders of magnitude. For
example, when $B_{12}$ increases from 10 to $10^3$, the rates
increase from 0.1749 to 25.5680 at $\rho_7=0.01, T_9=0.1$, and from
0.0022 to 3.8848 at $\rho_7=0.1, T_9=0.2$. The considerable increase
in the reaction rates for $^{23}$Mg $(p, \gamma)$ $^{24}$Al implies
that more $^{23}$Mg will escape the Ne-Na cycle due to SES in a SMF.
Then it will make the next reaction convert more $^{24}$Al
$(\beta^+, \nu)$ $^{24}$Mg to participate in the Mg-Al cycle. It may
lead to synthesizing a large amount of heavy elements at the crust
of magnetars. These heavy elements, which are produced from the
nucleosynthesis process, may be thrown out due to the compact binary
mergers of double neutron star (NS-NS) or black hole and neutron
star (BH and NS) systems. On the other hand, our model for the rates
is in good agreement with those of LD and FGP models at relatively
low density (e.g., $\rho_7<0.01$) and $B_{12}<10^2$. In relatively
low magnetic fields (e.g., $B_{12}<1$), the SES of LD and FGP models
have strong influence on the rates compare to our model. However,
the rates in our model can be about 1.58 times and three orders
magnitude higher than those of FGP and LD in relatively high
magnetic fields and low density surroundings (e.g.,
$B_{12}\geq10^2$, $\rho_7<0.05$), respectively. The results we
derived, may have very important implications in some astrophysical
applications for the nuclear reaction, the thermal evolution, and
numerical simulation of magnetars.

\acknowledgments

We would like to thank the anonymous referee for carefully reading
the manuscript and providing some constructive suggestions which are
very helpful to improve this manuscript. This work was supported in
part by the National Natural Science Foundation of China under
grants 11565020, and the Counterpart Foundation of Sanya under grant
2016PT43, the Special Foundation of Science and Technology
Cooperation for Advanced Academy and Regional of Sanya under grant
2016YD28, the Scientific Research Starting Foundation for 515
Talented Project of Hainan Tropical Ocean University under grant
RHDRC201701, and the Natural Science Foundation of Hainan Province
under grant 114012.

\clearpage

\begin{figure*}
\centering
\includegraphics[width=10cm,height=10cm]{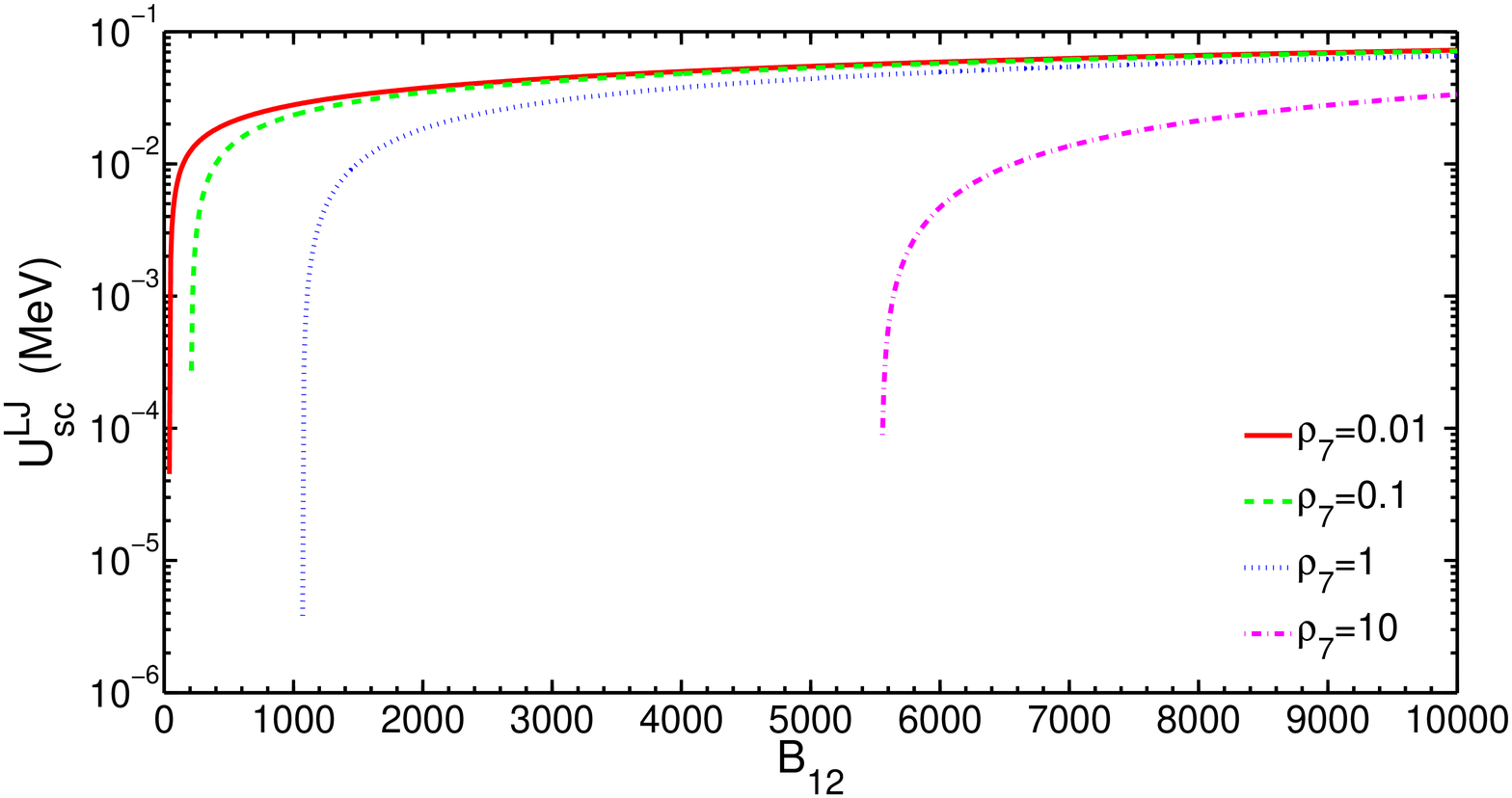}
  \caption{The electron screening potential as a function of $B_{12}$ of LJ model for some typical astronomical condition.\label{fig1}}
\end{figure*}
\begin{figure*}
\centering
    \includegraphics[width=6cm,height=6cm]{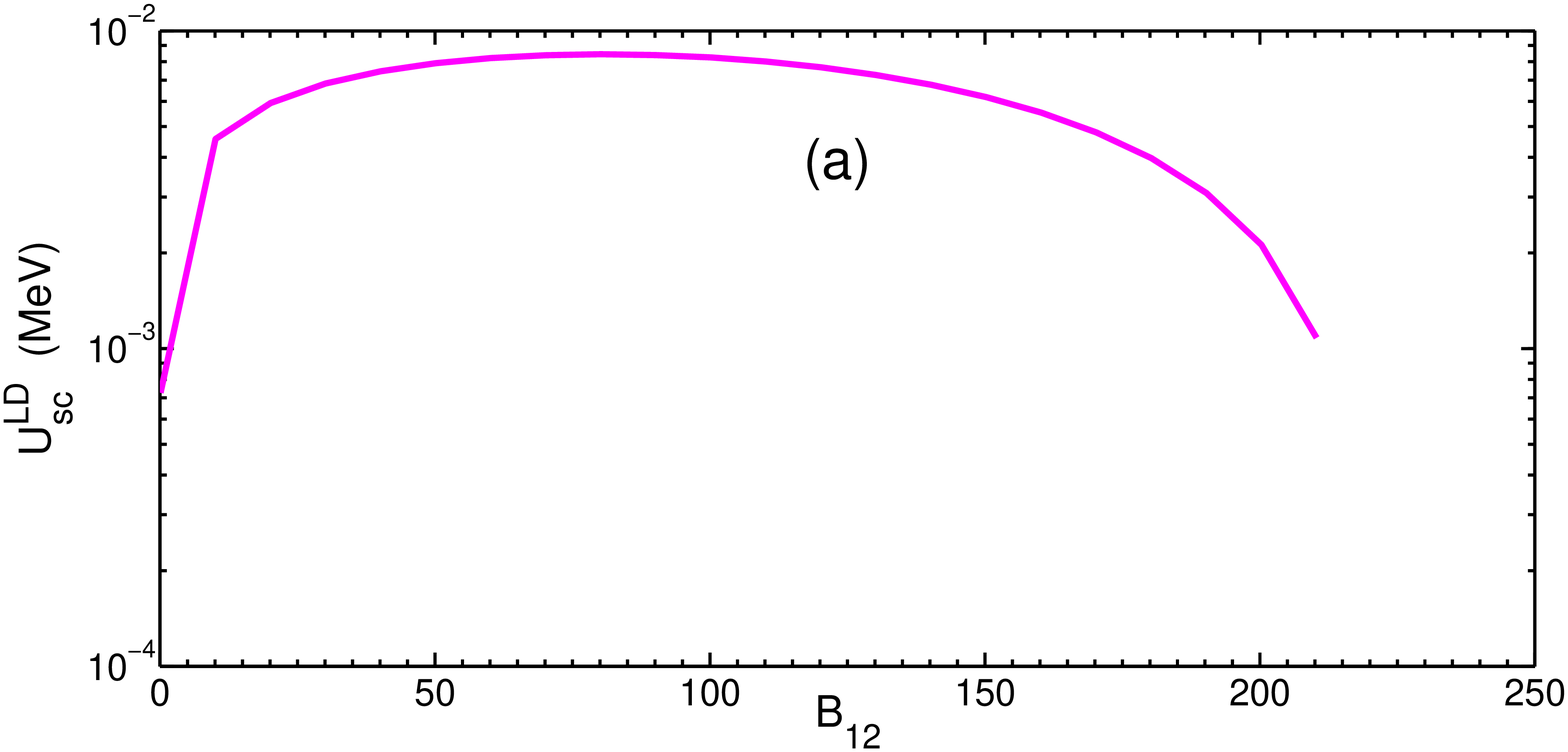}
    \includegraphics[width=6cm,height=6cm]{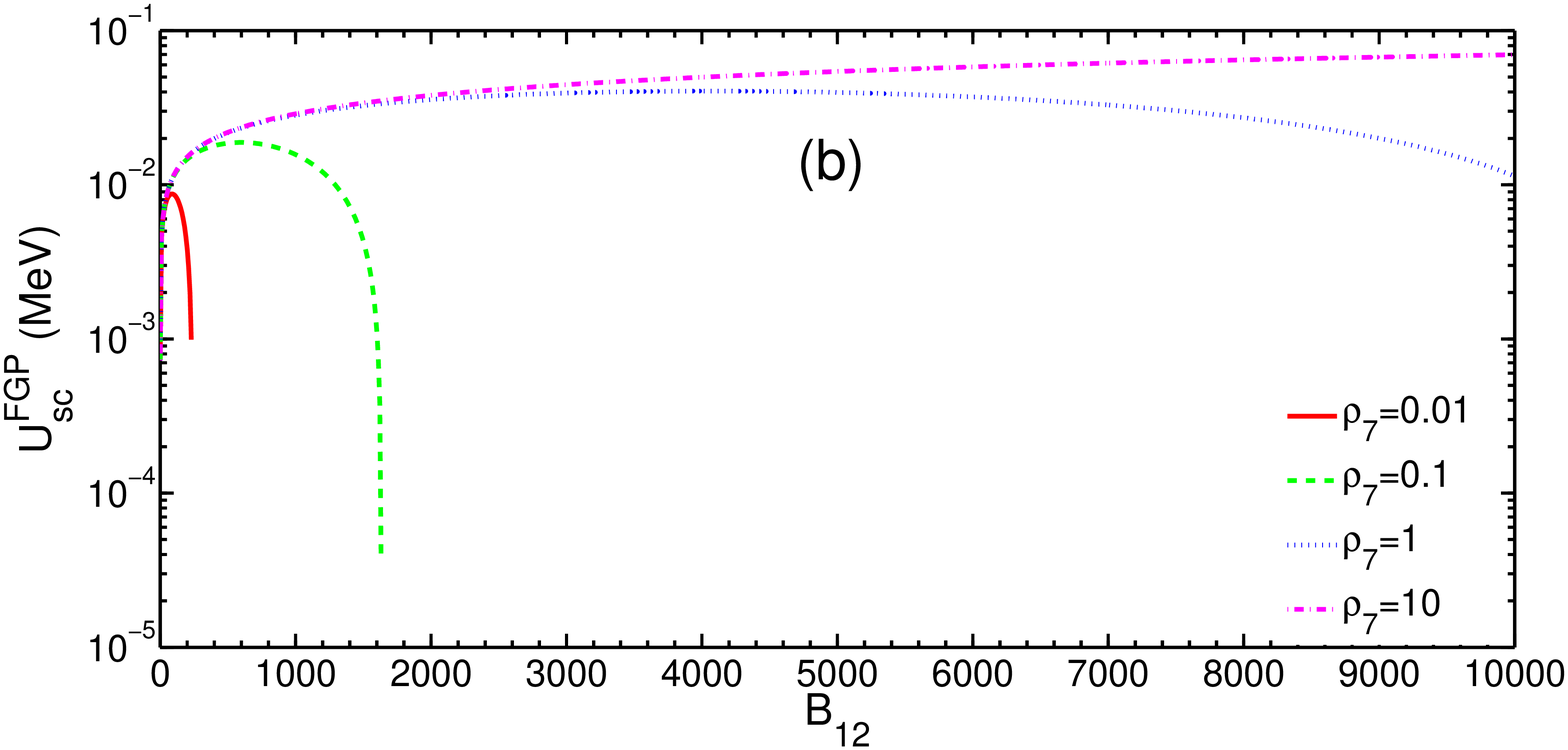}
\caption{The electron screening potential as a function of $B_{12}$
in LD, and FGP models for some typical astronomical
condition.\label{fig2}}
\end{figure*}
\begin{figure*}
  \centering
\includegraphics[width=10cm,height=10cm]{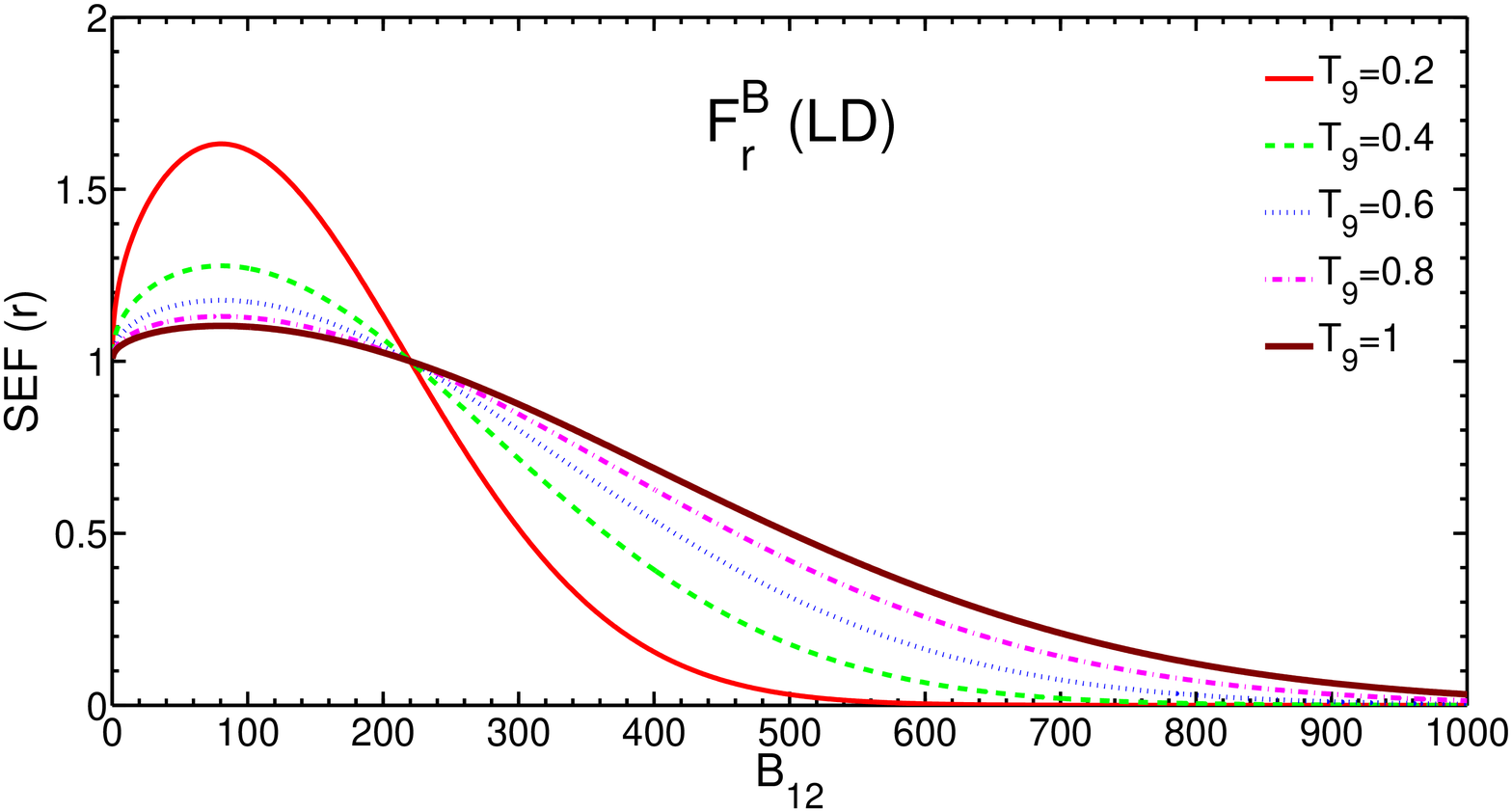}
  \caption{The resonant SEF for LD model as a function of $B_{12}$
in the case with SES and SMF.\label{fig3}}
\end{figure*}

\begin{figure*}
\centering
\includegraphics[width=6cm,height=6cm]{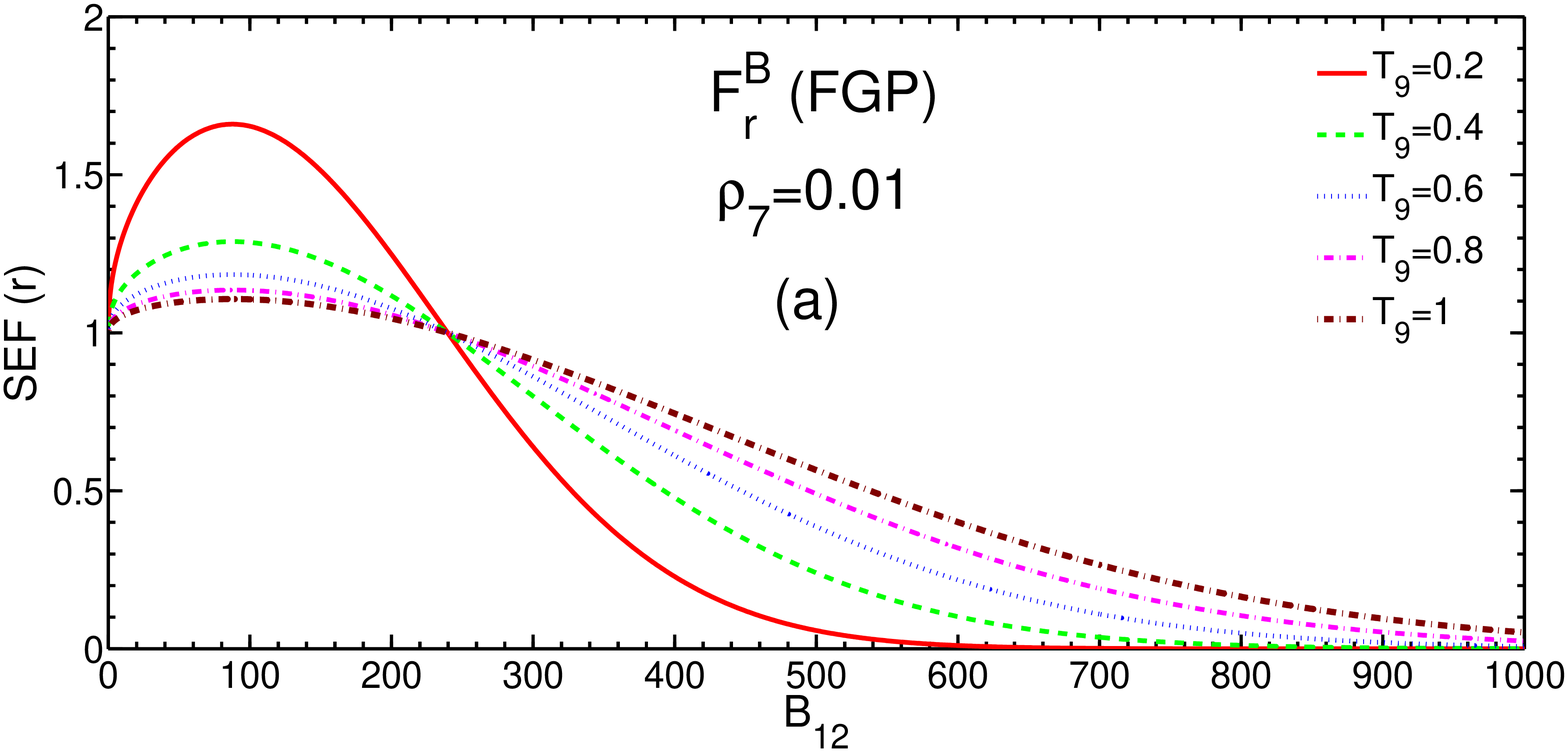}
    \includegraphics[width=6cm,height=6cm]{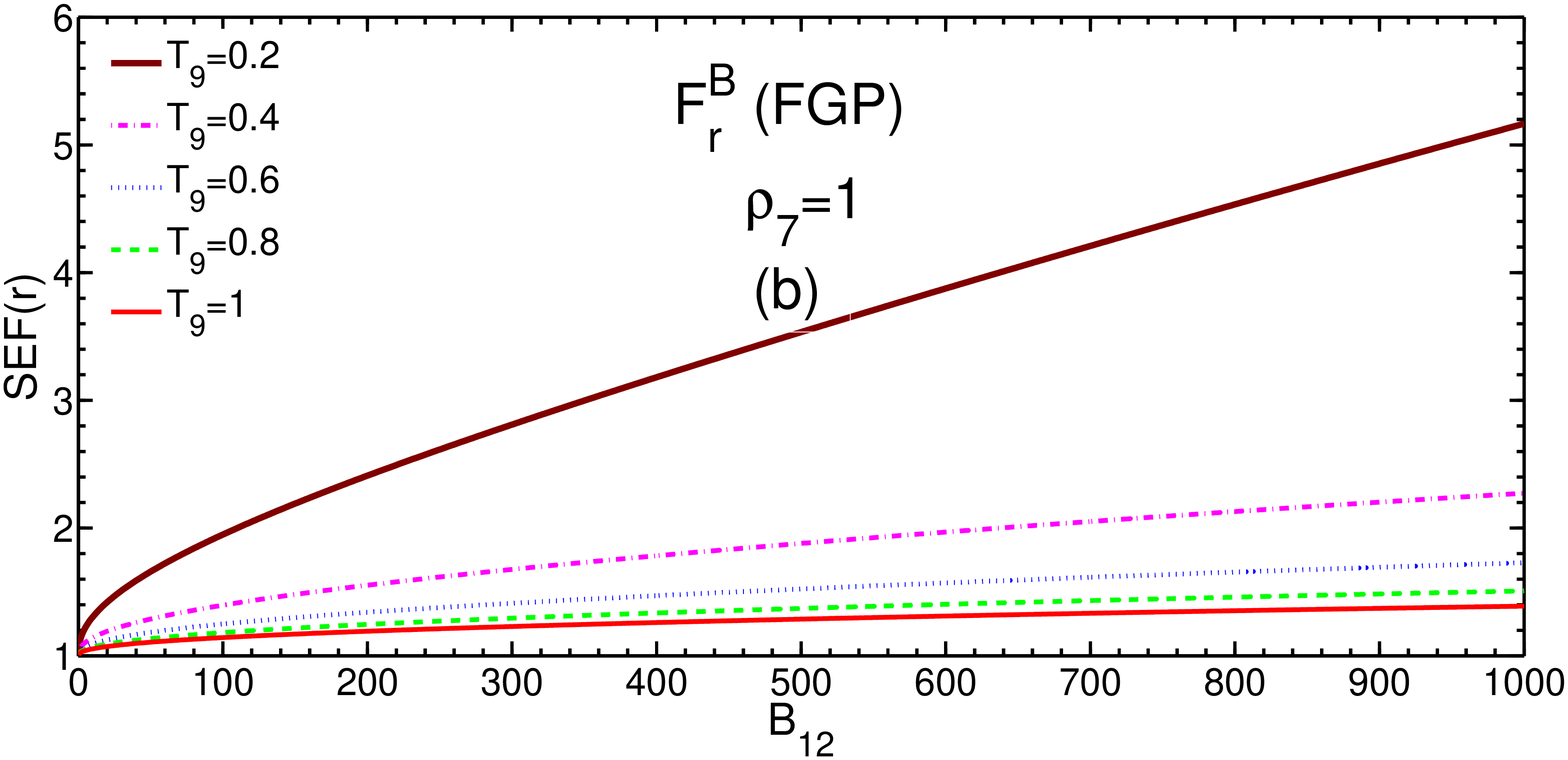}
    \includegraphics[width=6cm,height=6cm]{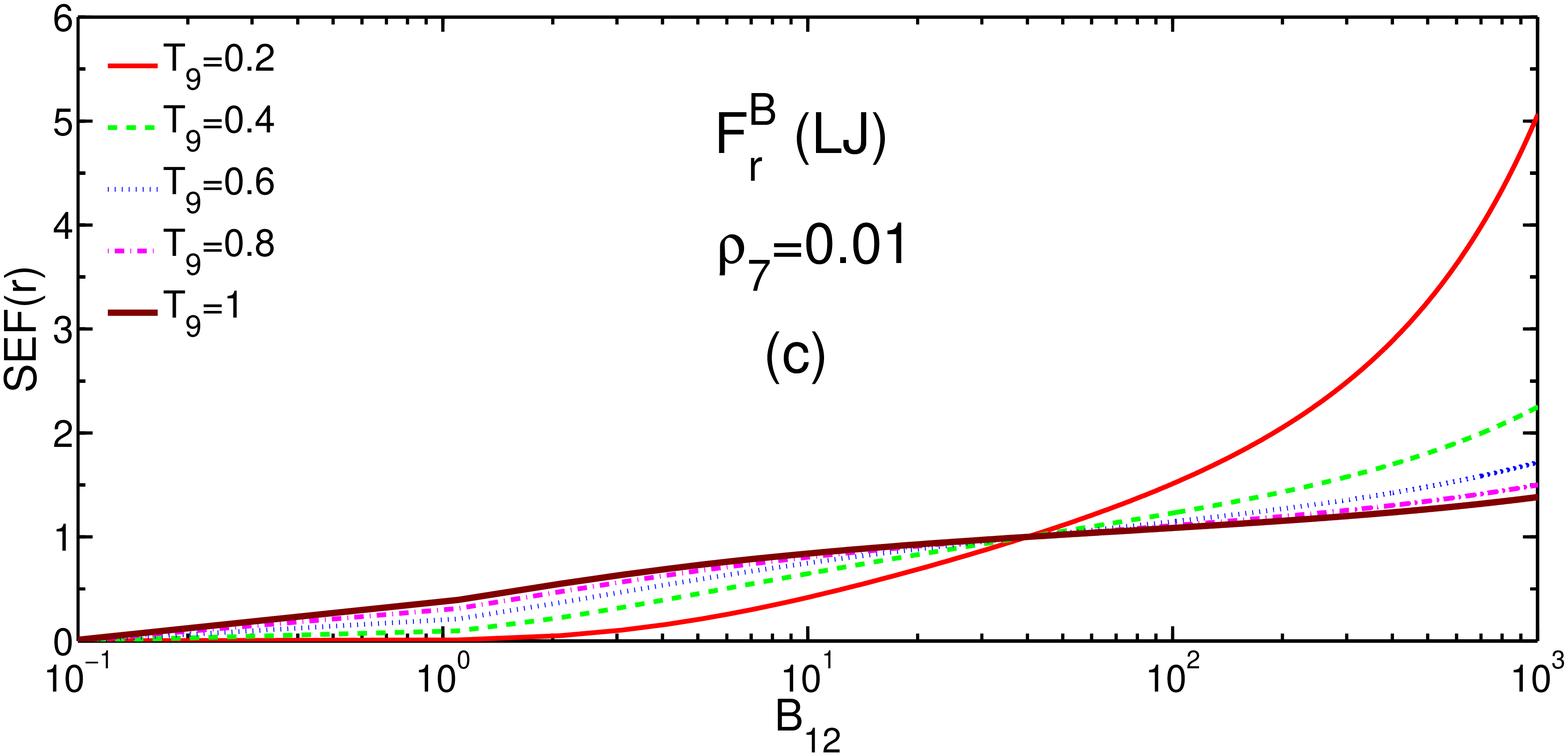}
    \includegraphics[width=6cm,height=6cm]{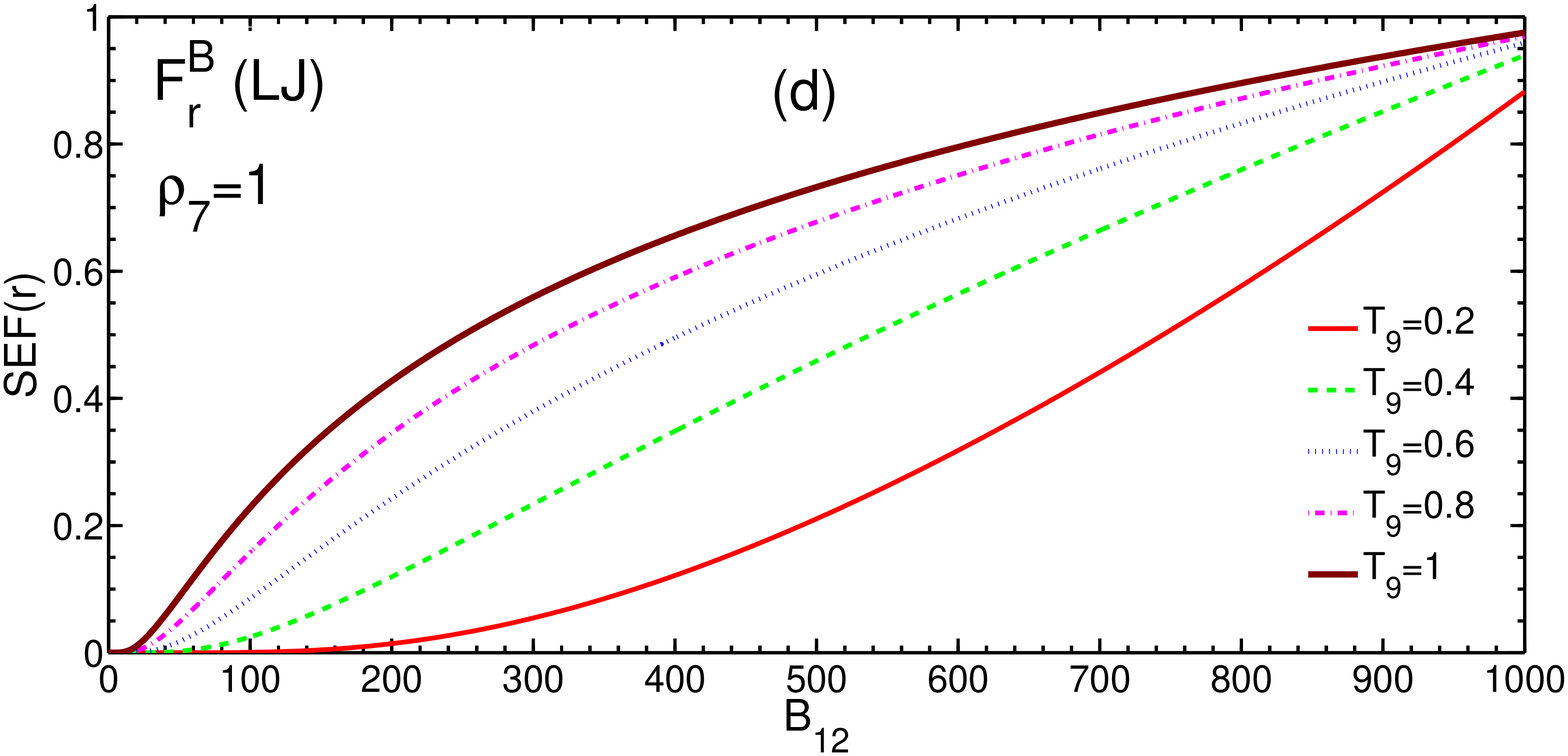}

\caption{The resonant SEF for FGP and LJ models as a function of
$B_{12}$ in the case with SES and SMF.\label{fig4}}
\end{figure*}


\begin{figure*}
\centering
     \includegraphics[width=6cm,height=6cm]{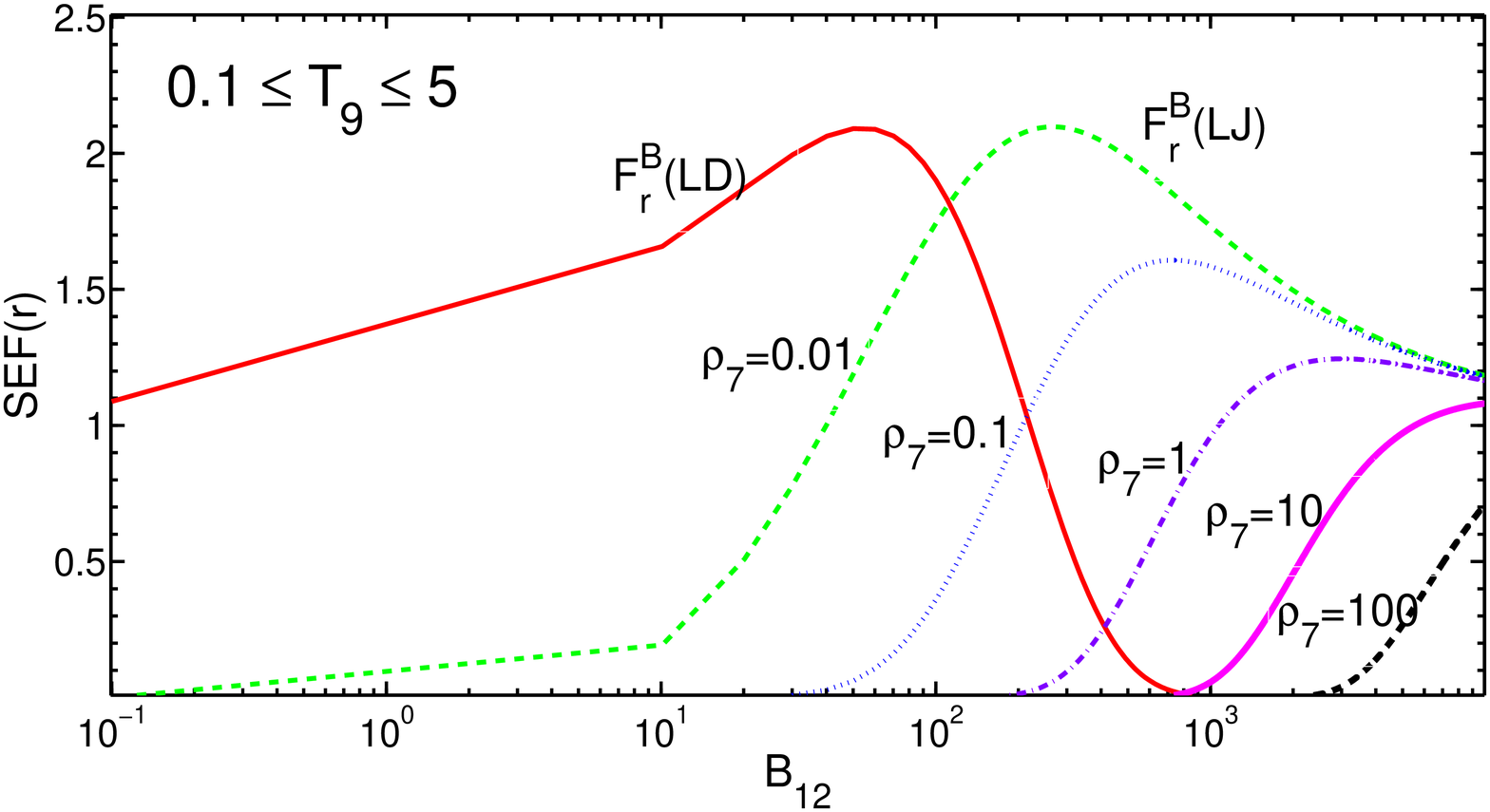}
    \includegraphics[width=6cm,height=6cm]{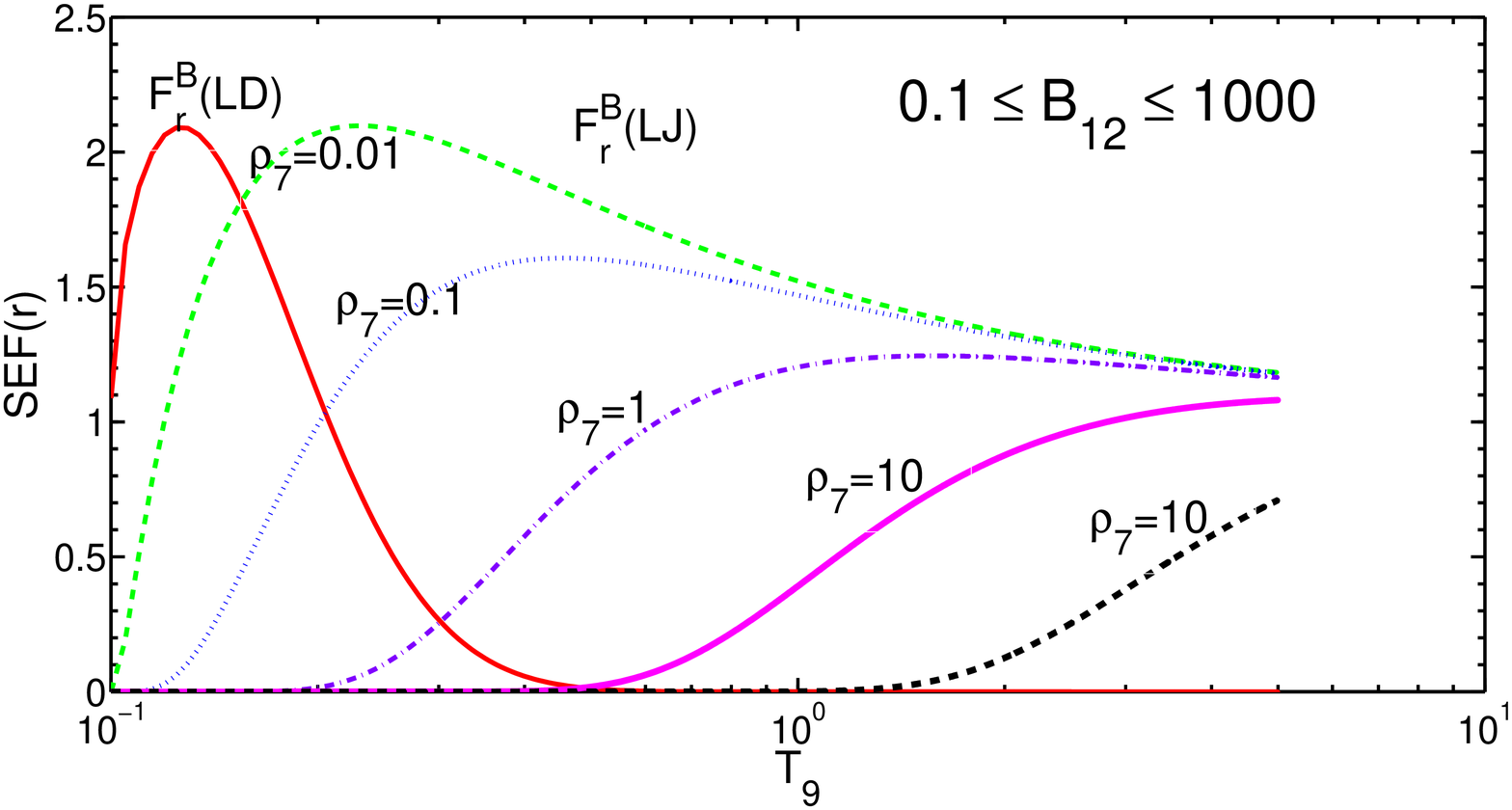}
    \includegraphics[width=6cm,height=6cm]{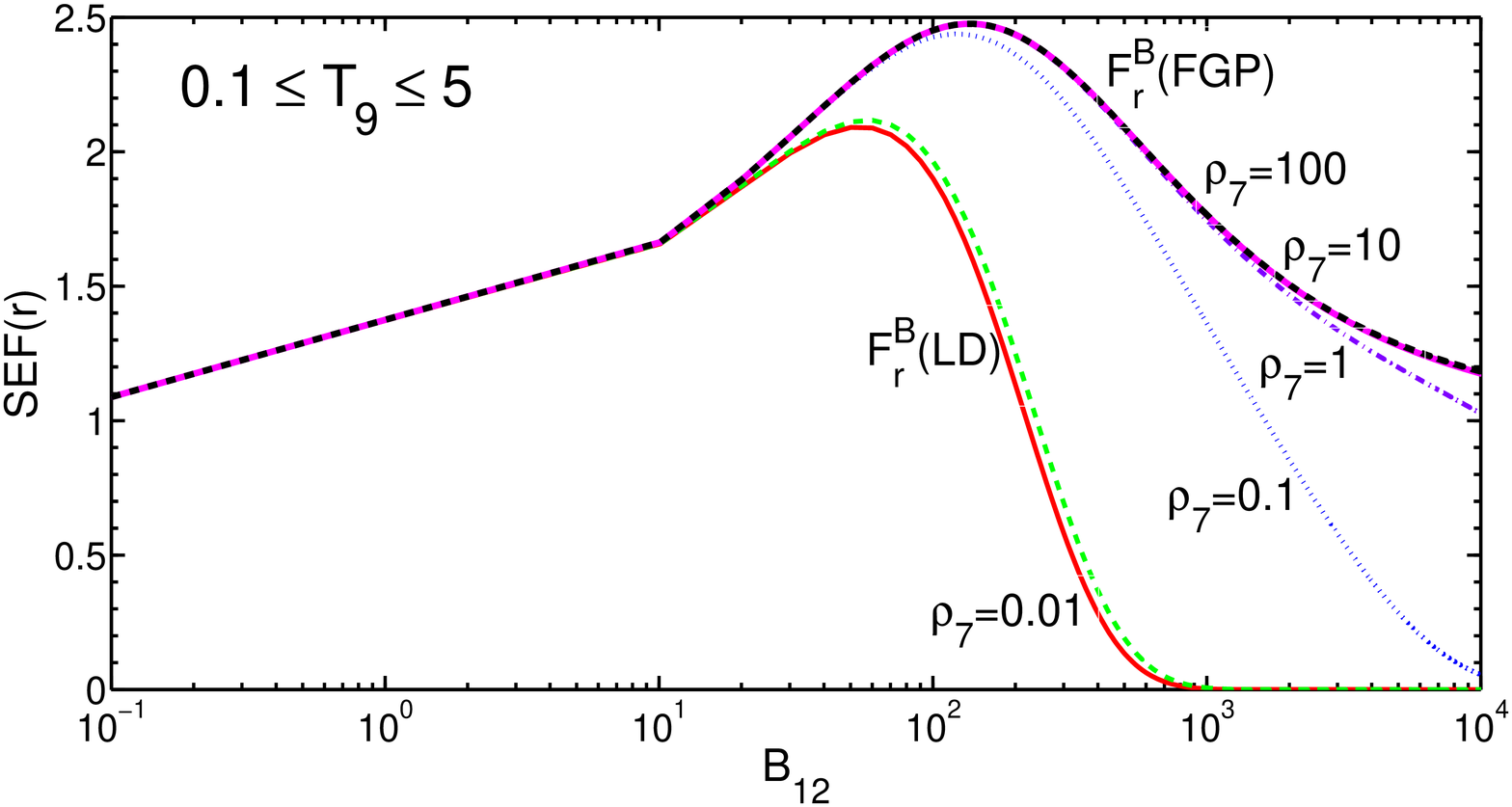}
    \includegraphics[width=6cm,height=6cm]{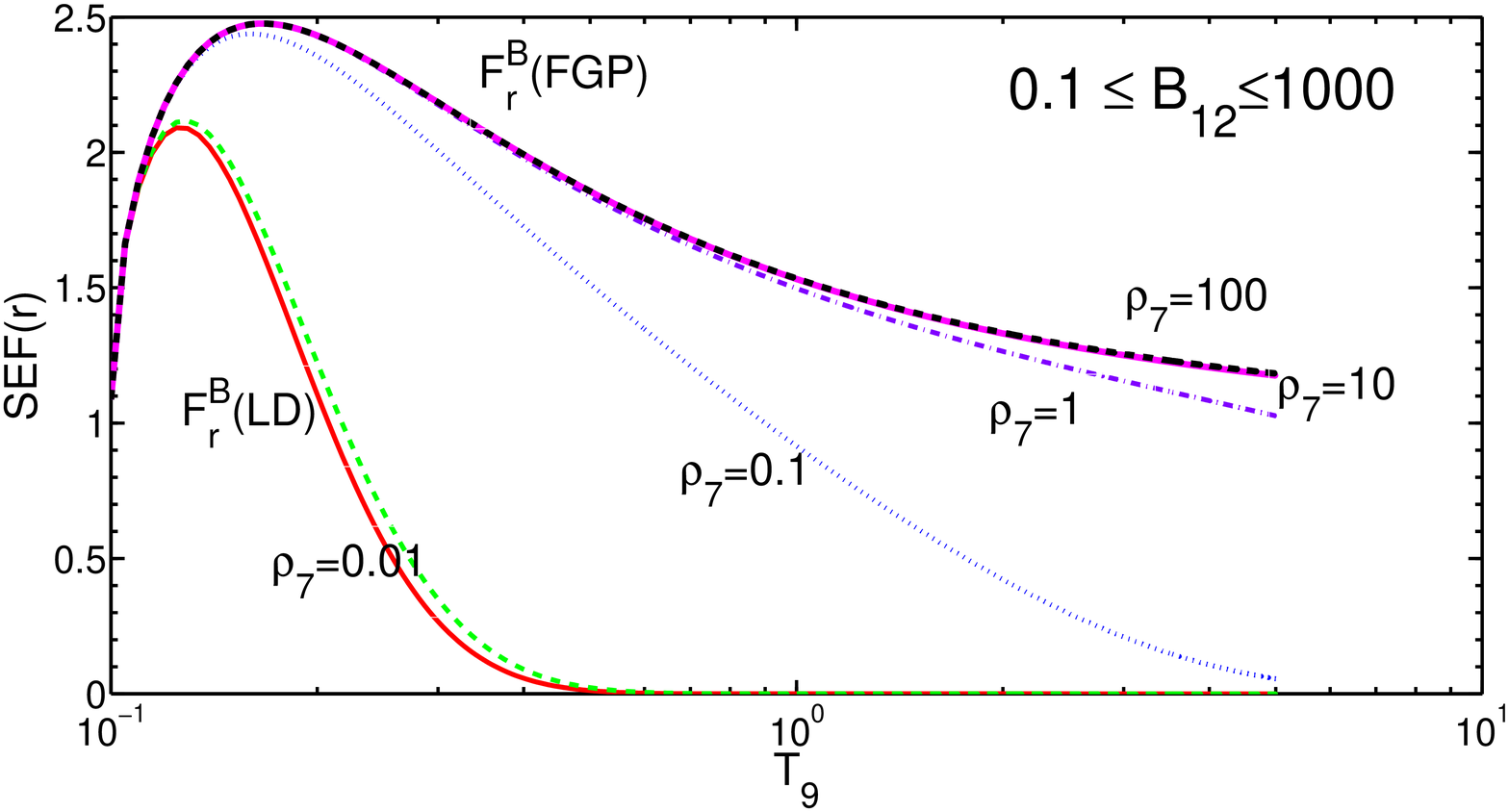}
    \includegraphics[width=6cm,height=6cm]{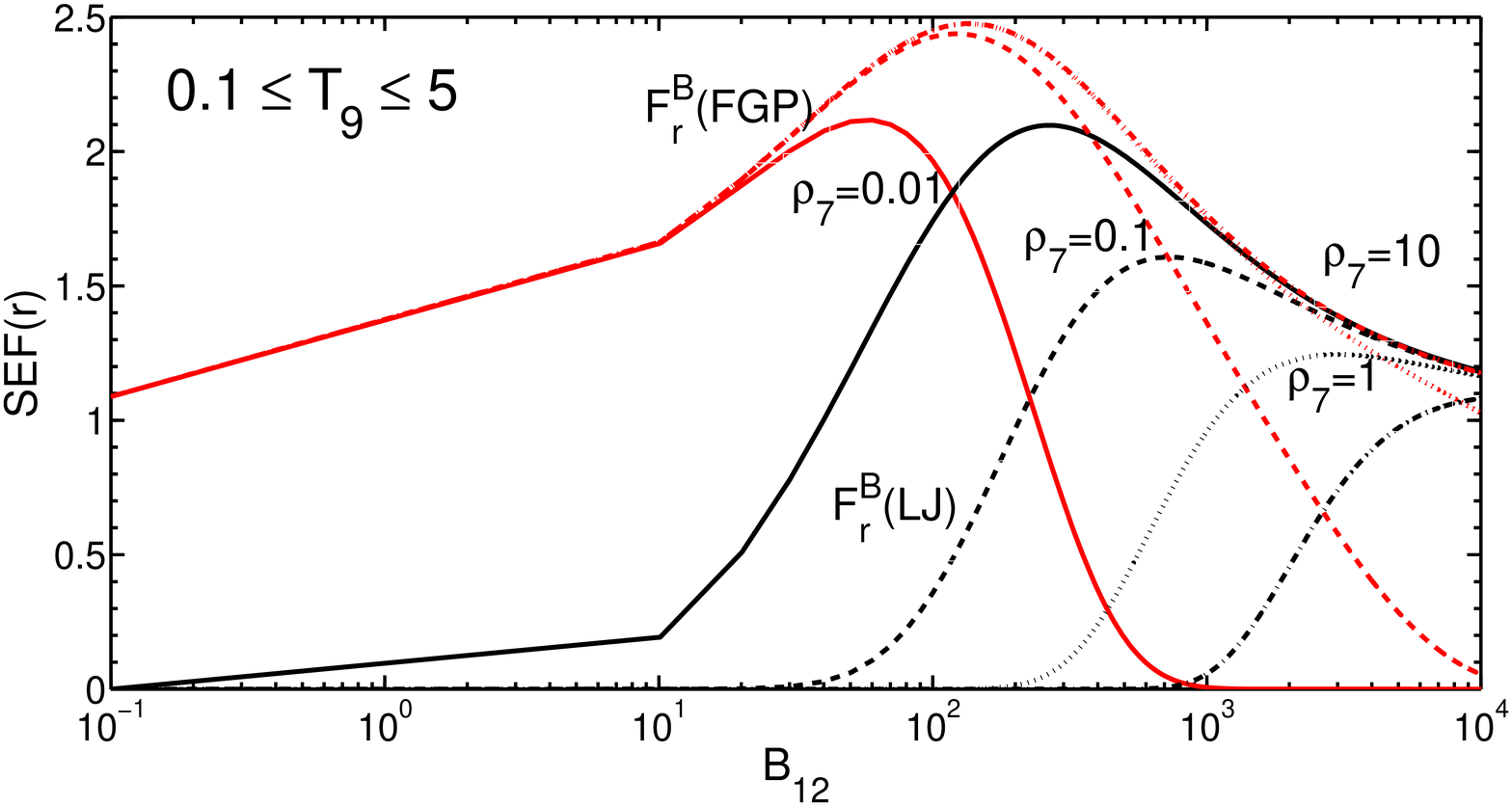}
    \includegraphics[width=6cm,height=6cm]{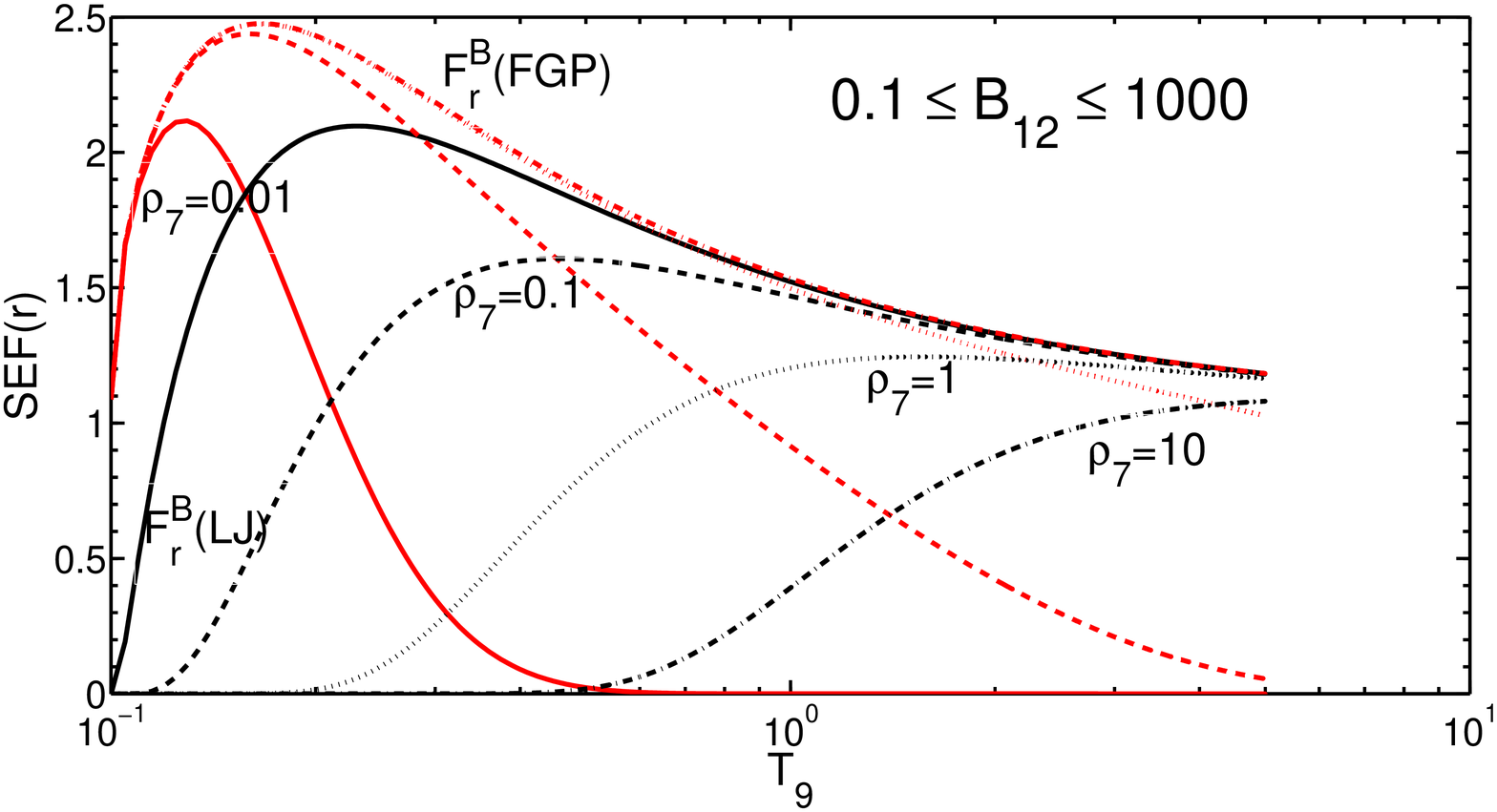}

\caption{The comparisons are plotted for some typical astronomical
condition of the resonant SEF among the three models of LJ, LD, and
FGP in the case with SES and SMF.\label{fig5}}
\end{figure*}


\begin{figure*}
\centering
    \includegraphics[width=6cm,height=6cm]{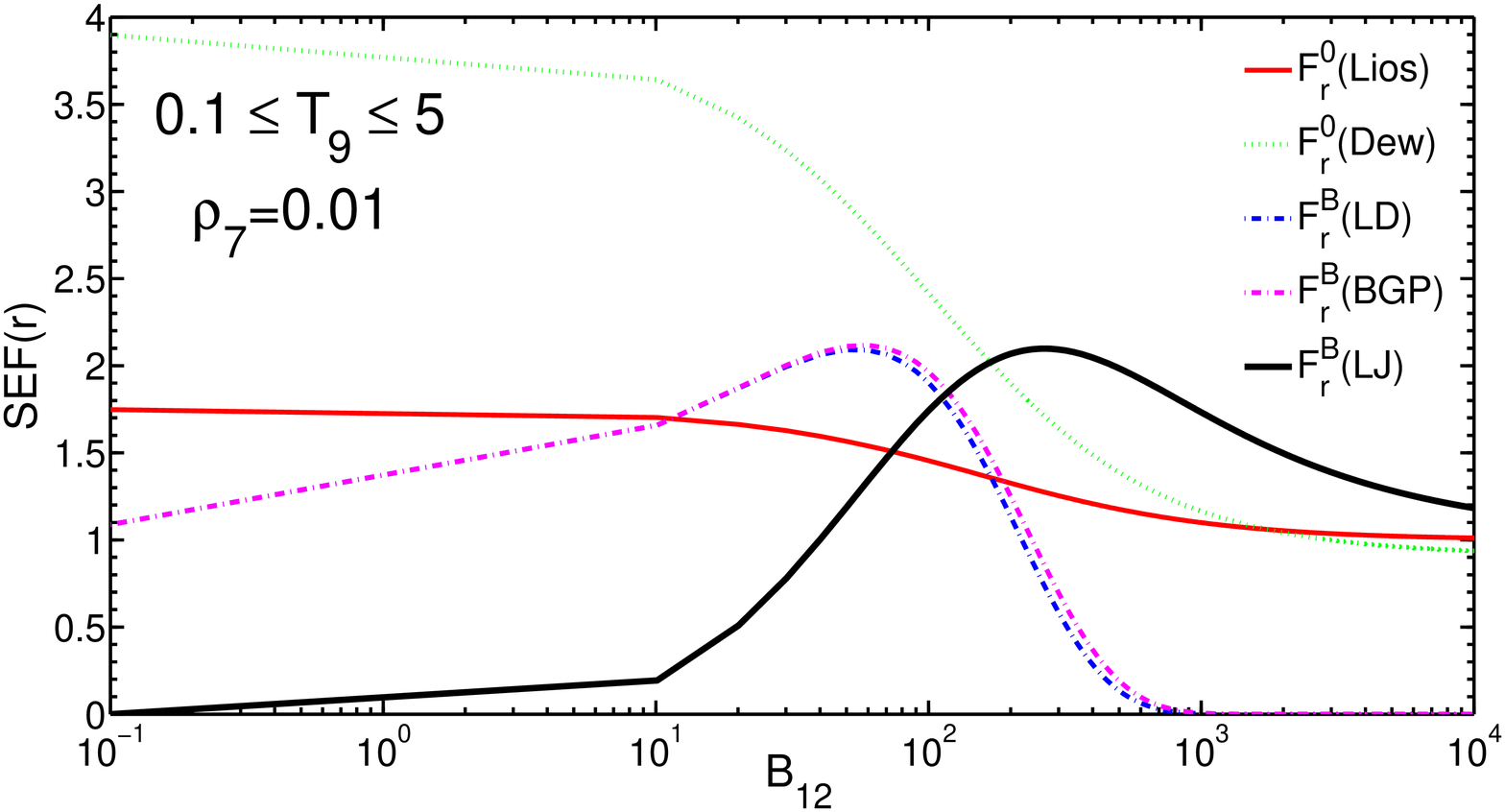}
    \includegraphics[width=6cm,height=6cm]{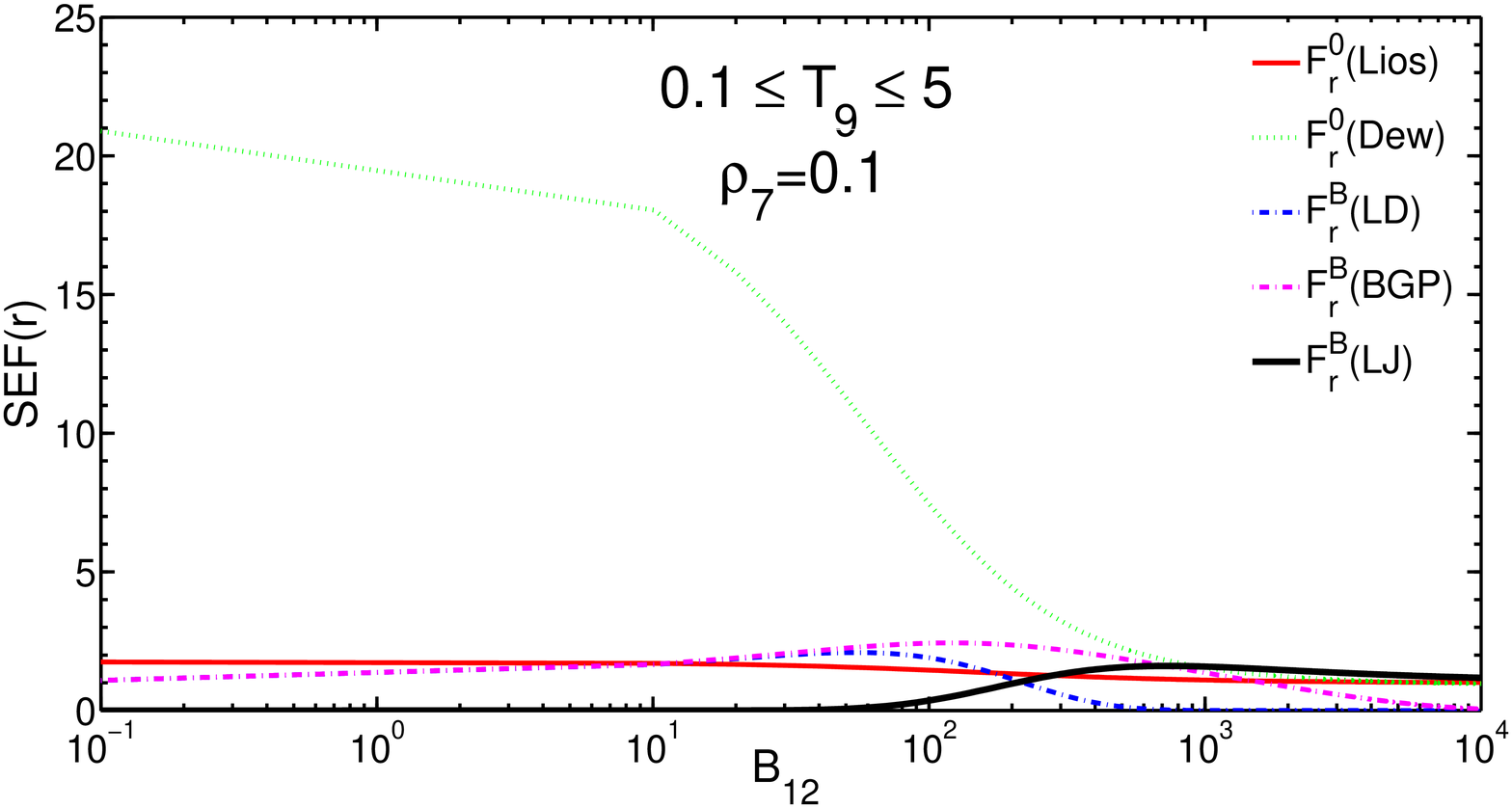}\\
    \includegraphics[width=6cm,height=6cm]{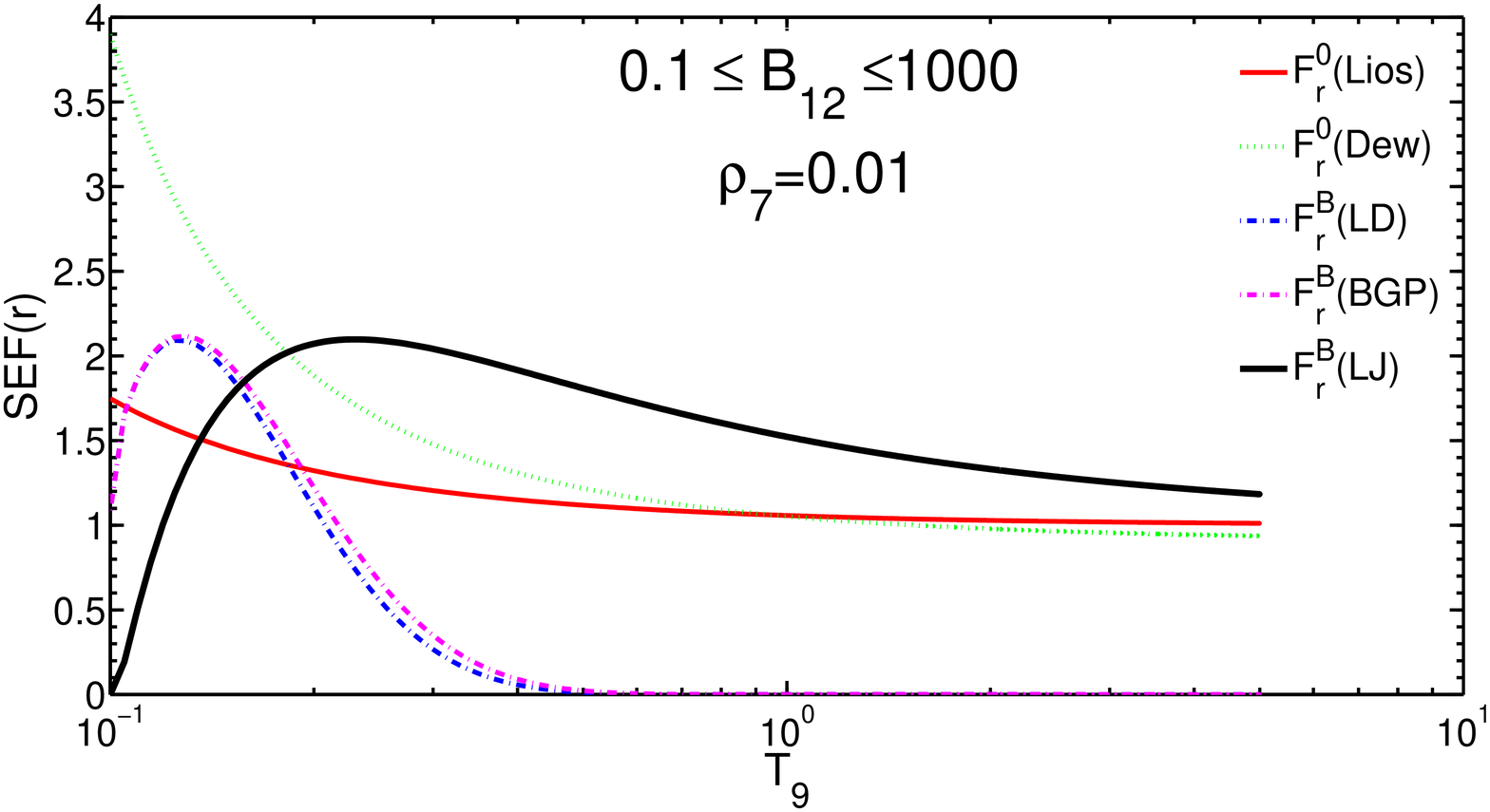}
    \includegraphics[width=6cm,height=6cm]{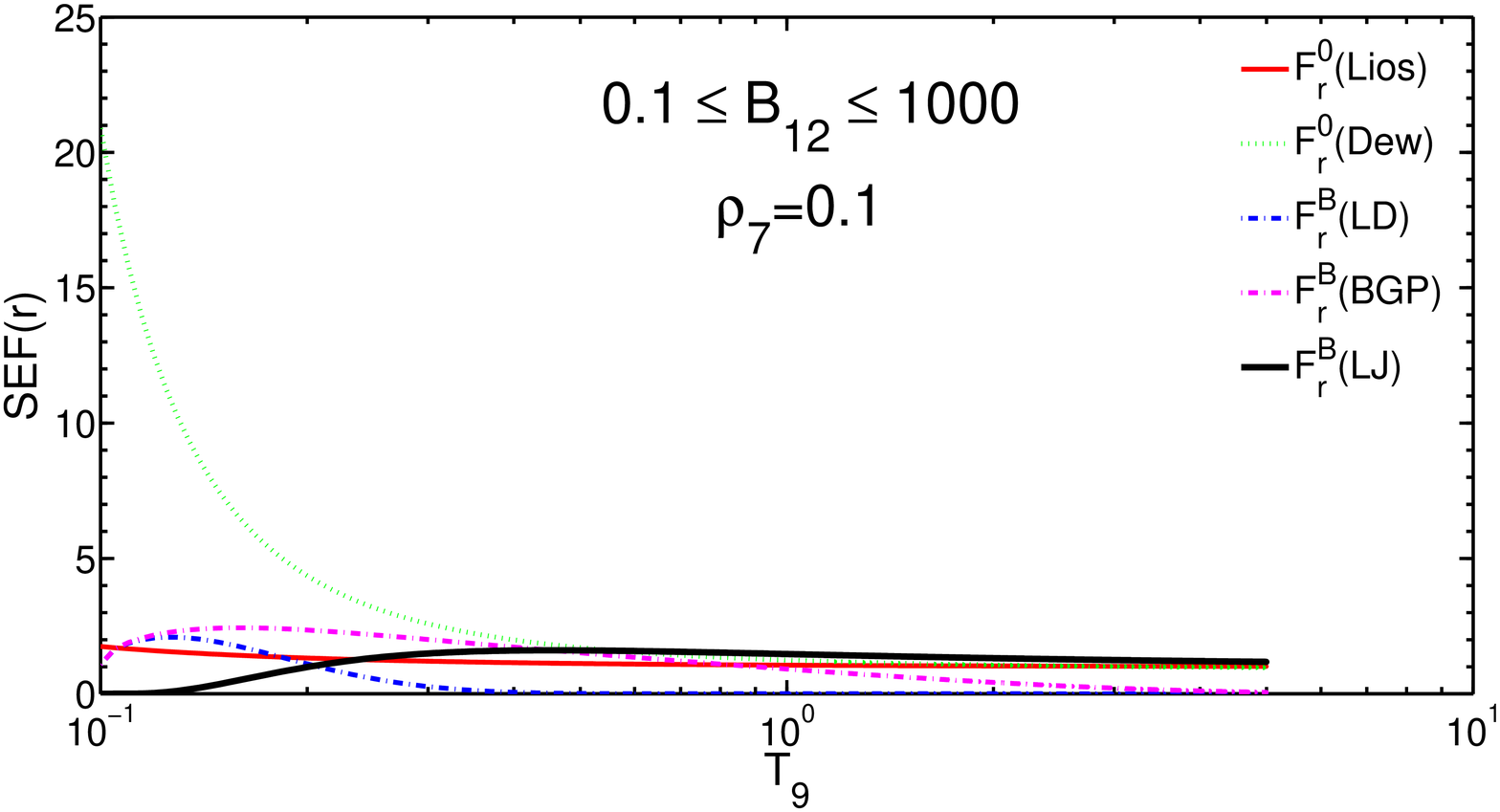}

 \caption{The comparisons of the resonant SEF for the model of Liolios, Dewitt with those of
models of LD, FGP, and LJ.\label{fig6}}
\end{figure*}

\end{document}